\documentclass[preprint]{iucr}              % DO NOT DELETE THIS LINE
\RequirePackage{graphicx}
\usepackage{amsmath}
\usepackage{mathtools}
\usepackage{cases}
\usepackage{upgreek}
     \journalcode{J}              % Indicate the journal to which submitted
\begin{document}                  % DO NOT DELETE THIS LINE

     %-------------------------------------------------------------------------
     % The introductory (header) part of the paper
     %-------------------------------------------------------------------------

     % The title of the paper. Use \shorttitle to indicate an abbreviated title
     % for use in running heads (you will need to uncomment it).

\title{A finite element approach to dynamical diffraction problems in reflection geometry}
%\shorttitle{Short Title}

     % Authors' names and addresses. Use \cauthor for the main (contact) author.
     % Use \author for all other authors. Use \aff for authors' affiliations.
     % Use lower-case letters in square brackets to link authors to their
     % affiliations; if there is only one affiliation address, remove the [a].

\author[a]{Ari-Pekka}{Honkanen}
\author[b]{Claudio}{Ferrero}
\author[b]{Jean-Pierre}{Guigay}
\cauthor[c]{Vito}{Mocella}{vito.mocella@cnr.it}{}

\aff[a]{University of Helsinki, Department of Physics, PO Box 64, FI-00014 Helsinki, \country{Finland}}
\aff[b]{ESRF - The European Synchrotron, Grenoble, \country{France}}
\aff[c]{CNR-IMM Sede di Napoli, via P. Castellino 111, I-80131 Napoli, \country{Italy}}

     % Use \shortauthor to indicate an abbreviated author list for use in
     % running heads (you will need to uncomment it).

%\shortauthor{Soape, Author and Doe}

     % Use \vita if required to give biographical details (for authors of
     % invited review papers only). Uncomment it.

%\vita{Author's biography}

     % Keywords (required for Journal of Synchrotron Radiation only)
     % Use the \keyword macro for each word or phrase, e.g. 
     % \keyword{X-ray diffraction}\keyword{muscle}

%\keyword{keyword}

     % PDB and NDB reference codes for structures referenced in the article and
     % deposited with the Protein Data Bank and Nucleic Acids Database (Acta
     % Crystallographica Section D). Repeat for each separate structure e.g
     % \PDBref[dethiobiotin synthetase]{1byi} \NDBref[d(G$_4$CGC$_4$)]{ad0002}

%\PDBref[optional name]{refcode}
%\NDBref[optional name]{refcode}

\maketitle                        % DO NOT DELETE THIS LINE

\begin{synopsis}
Supply a synopsis of the paper for inclusion in the Table of Contents.
\end{synopsis}

\begin{abstract}
A finite element approach to solve numerically the Takagi-Taupin equations expressed in a weak form is presented and applied to simulate X-ray reflectivity curves, spatial intensity distributions and focusing properties of bent perfect crystals in symmetric reflection geometry. The proposed framework encompasses a new formulation of the Takagi-Taupin equations, which appears to be promising in terms of robustness and stability and supports the Fresnel propagation of the diffracted waves. The presented method is very flexible and has the potential of dealing with dynamical X-ray or neutron diffraction problems related to crystals of arbitrary shapes and deformations. The reference implementation based on the commercial COMSOL Multiphysics\textregistered \ software package is available to the relevant user community. 
\end{abstract}

     %-------------------------------------------------------------------------
     % The main body of the paper
     %-------------------------------------------------------------------------
     % Now enter the text of the document in multiple \section's, \subsection's
     % and \subsubsection's as required.

\section{Introduction}

The Takagi-Taupin equations (TTE) are partial differential equations (PDE) which describe the dynamical Bragg diffraction in a perfect or deformed crystal \cite{penning61,takagi62,taupin64,takagi69,authierbook,apolloni08}.  
Analytical solutions exist only for a few cases \cite{katagawa74,litzman74,chukhovskii78}. In general, one has to resort to numerical solution of the TTE. An approximate approach to solve diffraction curves of large crystals was introduced recently \cite{honkanen14a,honkanen16}, and an iterative method starting from an integral expression of the TTE and involving a series expansion is used by Yan \& Li \cite{yan14}. 

Traditionally, the TTE are solved \cite{authier68,balibar67,epelboin79,epelboin85,gronkowski91,carvalho93} using a finite difference (FD) scheme easily implementable on a Cartesian mesh, but not on an arbitrary (e.g. deformed) mesh. In principle FD could be implemented on curved crystal surfaces using reciprocity method \cite{carvalho93b} but it is yet to be done. Furthermore, the incident wave is usually considered to be either a plane wave referring to an infinitely distant point-source or a so-called “spherical wave” referring to a point-source located on the crystal surface, whereas the intermediate case of an arbitrary finite distance between the source and the crystal applies to many actual situations \cite{lagomarsino02}.

Conversely, a finite element method (FEM) based on a weak numerical form of the differential TTE can potentially deal very well with cases of any incident wave and a crystal of any shape. A great advantage of this approach is that FEM implementations \cite{reddybook,odenbook} used for engineering problems are readily available and can be applied to X-ray diffraction problems \cite{mocella03,mocella15,honkanen17}. One of the benefits of using FEM is that it supplies a great deal of flexibility in the selection of discretisation, both in the elements that may be used to discretise space and the so called basis and test functions. Smaller elements in a region where the gradient of the sought-after function is large could be easily used. Another considerable advantage of FEM is that its theory is well established, due to the close relationship between the numerical and the weak formulation of a PDE problem. In the present work, the FEM TTE solver is implemented on a commercial software package (\textsc{COMSOL Multiphysics}\textregistered , \texttt{http://www.comsol.com}) and the method is verified in the case of Bragg reflection from both a perfect and a cylindrically bent crystal plate. 
Bent crystals have frequently been used as focusing elements on X-ray or neutron beamlines both in reflection and transmission geometry, e.g. \cite{tolentino88,chukhovskii94,podorov01,mocella04,mocella08,nesterets08,sutter10,guigay16}. Similarly, the focusing properties of elliptical multilayers have also been studied \cite{guigay08,morawe08,osterhoff13}.

The structure of the paper is as follows. We will derive an alternative form of the TTE which is particularly suitable for the FEM method at hand in terms of stability and computational efficiency. The boundary conditions for the derived TTE are discussed and set in place for the reflection geometry. The propagation of the diffracted wavefield is examined in the context of Fresnel diffraction. The weak form of the TTE are derived and the details of the COMSOL implementation are discussed. Finally, the validity of the method is investigated through a chosen set of simulations. This work is further development to our previously published work \cite{honkanen17}.

\section{Takagi-Taupin equations \label{sec:TT}}

Let us consider a crystal in Bragg diffraction geometry in which the incident beam is represented by a $\sigma$-polarised\footnote{The $\pi$-polarisation case can be described similarly} monochromatic modulated plane wave of the form 
\begin{equation}
\psi_{inc}(\mathbf{r}) = E_{inc}(\mathbf{r}) \exp(i\mathbf{k}_0\cdot\mathbf{r}).
\end{equation}
The length of the wavevector $\mathbf{k}_0$ is $2 \pi / \lambda$, where $\lambda$ is the wavelength of the X-ray. The diffracted wave in vacuum can be written analogously 
\begin{equation}
\psi_{out}(\mathbf{r}) = E_{out}(\mathbf{r}) \exp(i\mathbf{k}_h\cdot\mathbf{r}),
\end{equation}
where $\mathbf{k}_h = \mathbf{k}_0 + \mathbf{h}$ with $\mathbf{h}$ being the reciprocal vector corresponding to the diffractive planes. 

In non-homogenous medium, the wavefield $\psi$ fulfils the general wave equation
\begin{equation}
\nabla^2 \psi + k^2 \left[ 1 + \chi(\mathbf{r})\right] \psi = 0, \label{eq:waveeq}
\end{equation}
to which the solution in the usual two-beam case is of the form
\begin{equation}\label{eq:wavefield}
\psi(\mathbf{r}) = E_{0}(\mathbf{r}) \exp(i\mathbf{k}_0\cdot\mathbf{r})
+ E_{h}(\mathbf{r}) \exp(i\mathbf{k}_h\cdot\mathbf{r}).
\end{equation}
For periodic, deformed medium, the susceptibility $\chi$ can be expanded in a Fourier-series-like manner followingly
\begin{equation}
\chi(\mathbf{r}) = \chi_0 + \chi_{\bar{h}} \exp(-i\mathbf{h}\cdot(\mathbf{r}-\mathbf{u}))
+ \chi_{h} \exp(i\mathbf{h}\cdot(\mathbf{r}-\mathbf{u})) + \ldots, \label{eq:chi_expand}
\end{equation}
where $\mathbf{u}$ is the displacement field. By multiplying Eqs.~\eqref{eq:chi_expand}
with \eqref{eq:wavefield} and retaining only the terms relevant to the two-beam case, we obtain
\begin{equation}
\chi(\mathbf{r}) \psi(\mathbf{r})
\approx \left[ \chi_0 E_0 + \chi_{\bar{h}} \exp(i\mathbf{h}\cdot\mathbf{u}) E_h \right]
\exp(i\mathbf{k}_0\cdot\mathbf{r}) +\left[ \chi_0 E_h + \chi_{h} \exp(-i\mathbf{h}\cdot\mathbf{u}) E_0 \right]
\exp(i\mathbf{k}_h\cdot\mathbf{r}). \label{eq:chipsi}
\end{equation}
Since $E_{0,h}$ are slowly varying comparing $\exp(i\mathbf{k}_{0,h}\cdot\mathbf{r})$,
their second order derivatives arising in $\nabla^2 \psi$ can be neglected. Hence the following approximation applies
\begin{equation}
\nabla^2 \psi \approx  \left[ 2 i \mathbf{k}_0 \cdot \nabla E_0 - \mathbf{k}_0^2 E_0 \right]\exp(i\mathbf{k}_0\cdot\mathbf{r})
+\left[ 2 i \mathbf{k}_h \cdot \nabla E_h - \mathbf{k}_h^2 E_h \right]\exp(i\mathbf{k}_h\cdot\mathbf{r}) \label{eq:nab2psi}
\end{equation}
By substituting \eqref{eq:chipsi} and \eqref{eq:nab2psi} to \eqref{eq:waveeq}, we obtain
\begin{subnumcases}{}
2 \mathbf{k}_0 \cdot \nabla E_0 =  i \left[ k^2 (1+\chi_0) - k^2_0 \right] E_0
+ i k^2 \chi_{\bar{h}} E_h \exp(i\mathbf{h}\cdot\mathbf{u}) 
\label{eq:TT_Ea} \\
2 \mathbf{k}_h \cdot \nabla E_h =  i \left[ k^2 (1+\chi_0) - k^2_h \right] E_h
+ i k^2 \chi_{h} E_0 \exp(-i\mathbf{h}\cdot\mathbf{u}) \label{eq:TT_Eb}
\end{subnumcases}
Equations~\eqref{eq:TT_Ea} and \eqref{eq:TT_Eb} can be simplified by noting that $k_0^2 = k^2$ and 
$(k_h^2-k_0^2)/2k_h \approx k-k_h \approx k \Delta \theta \sin 2 \theta_B$,
where $\theta_B$ is the Bragg angle and $\Delta \theta = \theta - \theta_B$,
$\theta$ being the glancing angle of the incident wavevector $\mathbf{k}_0$ 
on the diffracting Bragg planes.

It is convenient to consider $E_{0,h}(\mathbf{r})$ as functions $E_{0,h}(s_0,s_h)$ of oblique coordinates $s_0$ and $s_h$ along the directions of  $\mathbf{k}_0$ and
 $\mathbf{k}_h$, respectively. As shown in Appendix~\ref{app:identities}, for any function $F(s_0,s_h)$ with gradient $\nabla F$ it holds $\mathbf{k}_{0,h}\cdot \nabla F = k_{0,h} \partial_{0,h} F(s_0,s_h)$, where $\partial_{0,h}$ denotes the partial derivative with respect to $s_{0,h}$. Thus Equations~\eqref{eq:TT_Ea} and \eqref{eq:TT_Eb} become
\begin{subnumcases}{}
2 \partial_0  E_0 =  i k \chi_0 E_0
+ i k \chi_{\bar{h}} E_h \exp(i\mathbf{h}\cdot\mathbf{u}) 
\label{eq:TT_E2a} \\
2 \partial_h E_h =  i \left[ k \chi_0 + 2 (k-k_h) \right] E_h
+ i k \chi_{h} E_0 \exp(-i\mathbf{h}\cdot\mathbf{u}).  \label{eq:TT_E2b}
\end{subnumcases}
The  case  of $\pi$-polarisation  can  be  included  in  this  formalism  by  replacing  the  coefficients $\chi_{\bar{h},h}$ by $C \chi_{\bar{h},h}$, where $C=1$ or $\cos 2\theta_B$ for $\sigma$- and $\pi$-polarisation, respectively. By using the notation
$c_{0}=k\chi_0/2$, $c_{\bar{h},h} = kC\chi_{\bar{h},h}/2$, $\beta = k-k_h \approx k \Delta \theta \sin 2 \theta_B$ and introducing the functions
\begin{align}
D_0(s_0,s_h)&=E_0(s_0,s_h)\exp\left(-ik\chi_0 \frac{s_0+s_h}{2} \right) \\
D_h(s_0,s_h)&=E_h(s_0,s_h)\exp\left(-ik\chi_0 \frac{s_0+s_h}{2} + i\mathbf{h}\cdot\mathbf{u}(s_0,s_h) \right),
\end{align}
Equations~\eqref{eq:TT_E2a} and \eqref{eq:TT_E2b} can be written as
\begin{subnumcases}{}
\partial_0  D_0 =  i c_{\bar{h}} D_h \label{eq:TT_typicala} \\
\partial_h  D_h =  i \left[ \beta + \partial_h (\mathbf{h}\cdot\mathbf{u}) \right] D_h
+ ic_h D_0. \label{eq:TT_typicalb}
\end{subnumcases}

Equations~\eqref{eq:TT_typicala} and \eqref{eq:TT_typicalb} are the most usual form of the TTE.
However, owing to the reasons explained in the next section, it is more convenient to use a modified expression in terms of the functions
\begin{align}
\Gamma_0 &= E_0 \exp(-i\beta s_h) \\
\Gamma_h &= E_h \exp(-i\beta s_h+i\mathbf{h}\cdot\mathbf{u})
\end{align}
By substituting the former to \eqref{eq:TT_E2a} and \eqref{eq:TT_E2b}, the TTE becomes
\begin{subnumcases}{}
\partial_0  \Gamma_0 =  i c_{0} \Gamma_0 + i c_{\bar{h}} \Gamma_h \label{eq:TT_gammaa} \\
\partial_h  \Gamma_h =  i \left[ c_0 + \partial_h (\mathbf{h}\cdot\mathbf{u}) \right] \Gamma_h + ic_h \Gamma_0  \label{eq:TT_gammab}
\end{subnumcases}

Equations~\eqref{eq:TT_gammaa} and \eqref{eq:TT_gammab} form the basis for our FEM implementation. The main advantage gained by moving the $\beta$ term out of the equation to the boundary conditions is the increased stability. The reason behind this can be understood by considering Eq.~\eqref{eq:TT_typicalb}. At the large $\beta$ limit,
the solution of $D_h$ is found to be proportional to 
$\exp(i\beta s_h)$, meaning that the phase of the solution oscillates rapidly along the propagation direction of the diffracted beam. In the length scale of the problem, these oscillations even out and thus have little physical consequence. However, they cause a major computational difficulty. This problem is well avoided by moving $\beta$ to the surface term, as $s_h$ varies slower along the surface than it would along the direction of the diffracted beam path. Thus a sparser solving grid can be used leading to shorter computation times and less heavy memory usage.

\section{Boundary conditions for the reflection geometry}

The handling of the TTE in Section~\ref{sec:TT} is valid for reflection, transmission, and mixed cases. The different cases are separated from each other \emph{via} the boundary conditions. For simplicity, we focus solely on the reflection geometry henceforth.

\begin{figure}
\label{fig:paraxial}
\centering
\includegraphics[width=0.5\textwidth]{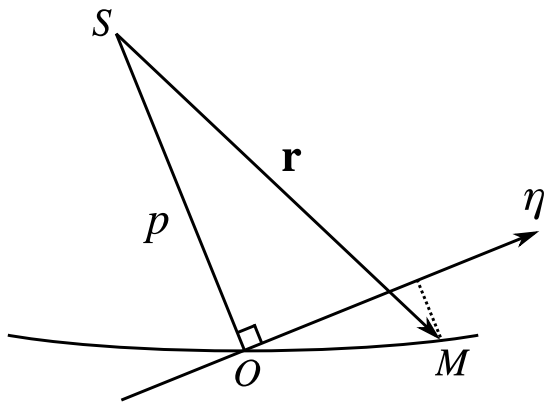}
\caption{The nomenclature used in the paraxial approximation.}
\end{figure}

\begin{figure}
\label{fig:boundaries}
\centering
\includegraphics[width=\textwidth]{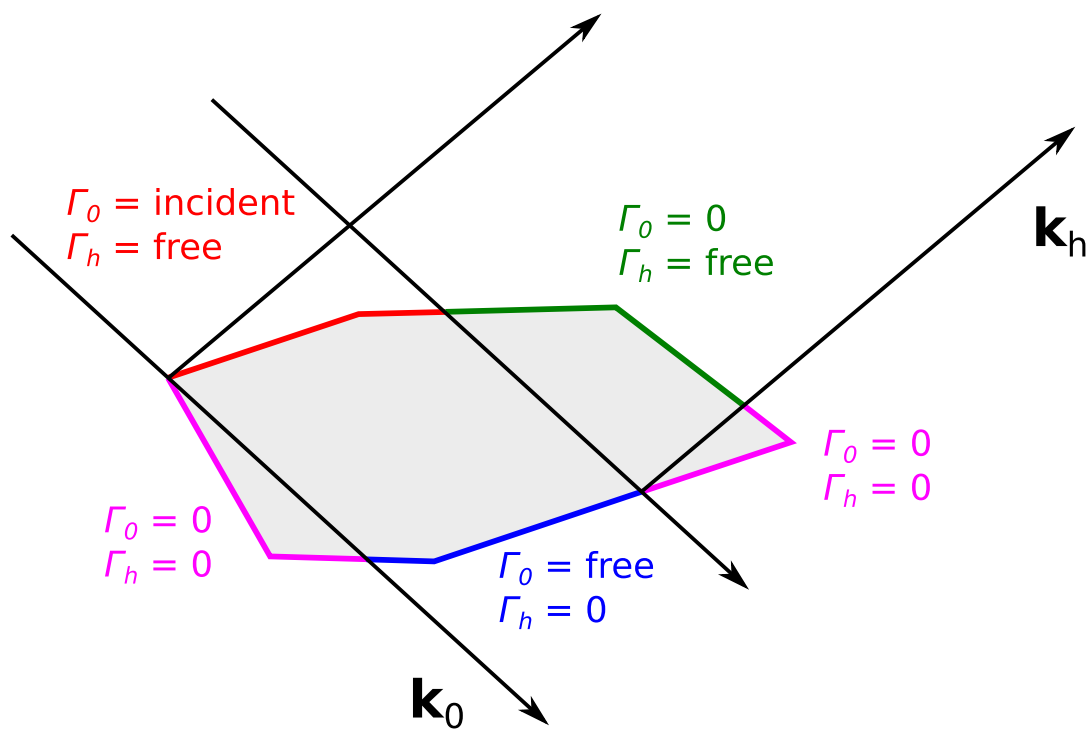}
\caption{Boundary conditions for the reflection geometry. Red: entrance surface of the incident wave; Blue: exit surface of the incident wave; Red + Green: exit surface of teh diffracted wave; Magenta: boundaries outside the domain of diffraction}
\end{figure}

On the entrance surface of the incident wave, the boundary condition for $\Gamma_0$ is
given by
\begin{equation}\label{eq:planewave}
\Gamma_0(\mathbf{r}_{surf}) = E_0 (\mathbf{r}_{surf})  \exp(-i\beta s_{h,surf}),
\end{equation}
where $s_{h,surf} = s_h(\mathbf{r}_{surf})$ is subject to the choice of the origin. For instance, for an incident plane wave $\psi_{inc,plane} = E_0 \exp(i\mathbf{k}_0\cdot\mathbf{r})$ 
\begin{equation}
\Gamma_{0,plane}(\mathbf{r}_{surf}) =  E_0 \exp(-i\beta s_{h,surf}),
\end{equation}
where $E_0$ is constant in this case. 
On the other hand, for a divergent source 
\begin{equation}
\psi_{inc,div} = \frac{A}{r^\gamma} \exp(i k r),
\end{equation}
where $\gamma = 1/2$ for a line source and $1$ for point source. Denoting as $\mathbf{r} = \overline{SM}$ the position vector of a point $M$ on the crystal surface with respect to the source $S$ and $\eta$ being the coordinate perpendicular to $\mathbf{k}_0$ (see Fig.~\ref{fig:paraxial}), we may use the so-called paraxial approximation
\begin{equation}
kr - \mathbf{k}_0\cdot\mathbf{r} \approx \frac{k \eta^2}{2r} \approx \frac{k \eta^2}{2 p},
\end{equation}
where $p$ is the distance from $S$ to the origin $O$ on the crystal surface such that the ray $\overline{SO}$ corresponds to the exact incident Bragg direction.
%By Taylor expanding the distance $r$ at the origin residing on the crystal surface, we find that $r \approx p + s_0 + \eta^2/2p$, where $p$ is the distance from the origin to the source and $\eta$ is the coordinate perpendicular to $s_0$. Dropping out the constant phase factor, we find $\psi_{inc,div} \approx A p^{-\gamma} \exp(i k s_0 + i k \eta^2/2p)$, so 
We thus obtain $\psi_{inc,div} \approx A p^{-\gamma} \exp(i \mathbf{k}_0\cdot\mathbf{r} + i k \eta^2/2p)$, so  the boundary condition becomes \begin{equation}\label{eq:pointsourcewave}
\Gamma_{0,div}(\mathbf{r}_{surf}) =  A p^{-\gamma} \exp\left(i k \frac{\eta^2}{2p} -i\beta s_h\right).
\end{equation}
In addition, the boundary condition for $\Gamma_0$ is left free (\emph{i.e.} to be solved) on the exit surface of the incident wave and set zero elsewhere (Fig.~\ref{fig:boundaries}).

For the diffracted wave $\Gamma_h$, the boundary condition is $\Gamma_h=0$ everywhere else except on the exit surface, where it is left free. One should note that the different surfaces may overlap with each other. The different boundaries are illustrated in Figure~\ref{fig:boundaries}.  

\section{Propagation of the diffracted wave}

\begin{figure}
\label{fig:fresnel_integral}
\centering
\includegraphics[width=.8\textwidth]{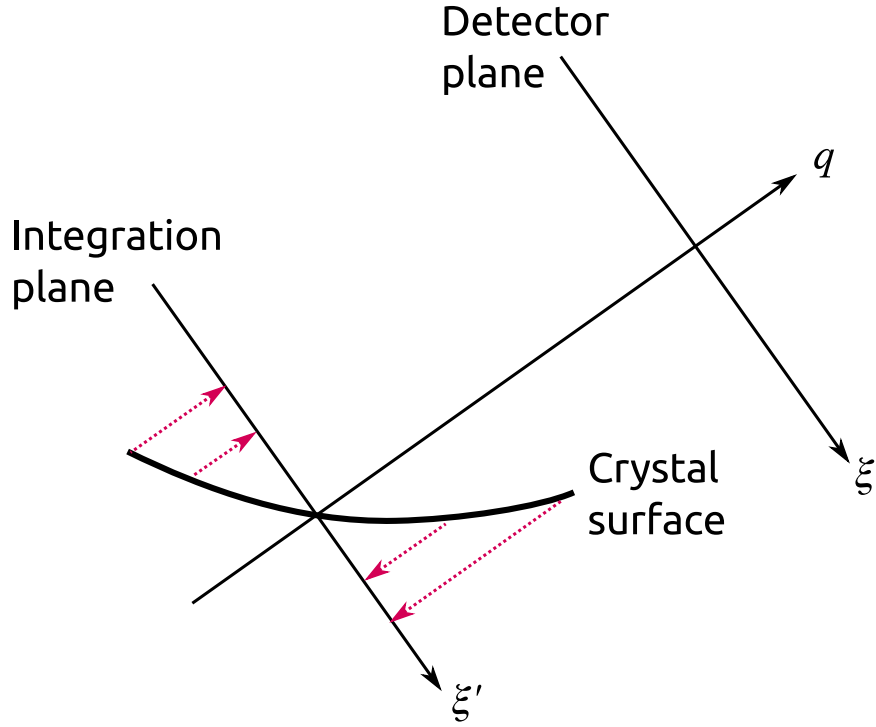}
\caption{Auxiliary planes and coordinate system for computing the wave propagation. In the vicinity of the crystal, the diffracted wave is propagated on the integration plane as a plane wave, from which it is further propagated to the detector plane using the Fresnel diffraction integral.}
\end{figure}

In order to describe the
propagation of the reflected beam in air, we use the rectangular coordinates
$(\xi,q)$ as depicted in Figure~\ref{fig:fresnel_integral}. For the solved $\Gamma_h$, the diffracted wave on the crystal surface is obtained by
\begin{equation}
\psi_{out}(\mathbf{r}_{surf}) =
\Gamma_h(\mathbf{r}_{surf}) \exp(i\mathbf{k}_h\cdot\mathbf{r}_{surf}+i\beta s_{h,surf}-i\mathbf{h}\cdot\mathbf{u}_{surf}),
\end{equation}
with $s_{h,surf}=s_h(\mathbf{r}_{surf})$ and $\mathbf{u}_{surf}=\mathbf{u}(\mathbf{r}_{surf})$. Since the diffracted wave is in essence a modulated plane wave, we can propagate
it (in the mathematical sense) in the vicinity of the surface simply by adjusting it's phase by $\exp(ik\Delta s_h)$. 

Now, let's consider a plane that goes through the origin ($s_h=0$) and is perpendicular to $s_h$.
If we propagate $\psi_{out}$ from the crystal surface on this plane, as indicated in Fig.~\ref{fig:fresnel_integral}, we find out that
\begin{equation}\label{eq:psi_outplane}
\psi_{out,plane}(\xi) = \Gamma_h(\mathbf{r}_{surf}(\xi)) \exp( i\beta s_{h,surf}(\xi)-i\mathbf{h}\cdot\mathbf{u}_{surf}(\xi)),
\end{equation}
where $\mathbf{r}_{surf}(\xi)$, $s_{h,surf}(\xi)$, and $\mathbf{u}_{surf}(\xi)$ evaluated at $\mathbf{r}_{surf}$ with the same $\xi$-coordinate.

Equation~\eqref{eq:psi_outplane} allows using the Fresnel diffraction integral in order to compute the wave amplitude far away from 
the crystal. In a plane at the distance $q$ from the origin, the wave amplitude is
\begin{equation}\label{eq:fresnel_integral}
\Psi(\xi,q) = \frac{1}{\sqrt{\lambda q}} \int d \xi' \ \psi_{out,plane}(\xi')
\exp \left(\frac{ik (\xi-\xi')^2}{2q} \right).
\end{equation}

\section{Weak formulation of Takagi-Taupin equations}

Following the well-established FEM procedure, both sides of Equations~\eqref{eq:TT_gammaa} and \eqref{eq:TT_gammab} are multiplied by test functions $v_1(s_0,s_h)$ and $v_2(s_0,s_h)$, and are integrated over the domain $\Omega$ with the boundary $\delta \Omega$:
\begin{subnumcases}{}
\int_\Omega dV \  \left(\partial_0  \Gamma_0 -  i c_{0} \Gamma_0 - i c_{\bar{h}} \Gamma_h \right) v_1 = 0 \label{eq:integralTTa} \\
\int_\Omega dV \ \left( \partial_h  \Gamma_h -  i \left[ c_0 + \partial_h (\mathbf{h}\cdot\mathbf{u}) \right] \Gamma_h - ic_h \Gamma_0 \right)v_2   = 0 \label{eq:integralTTb}
\end{subnumcases}

Let $\mathbf{s}_0$ and $\mathbf{s}_h$ be the unit vectors along the directions of $\mathbf{k}_0$ and $\mathbf{k}_h$, respectively. According to Appendix~\ref{app:identities}, we can write
\begin{align}
v_1 \partial_0 \Gamma_0 &= \partial_0 (v_1 \Gamma_0) - \Gamma_0 \partial_0 v_1 = \nabla \cdot (v_1 \Gamma_0 \mathbf{s}_0) - \Gamma_0 \partial_0 v_1   \\
v_2 \partial_h \Gamma_h &= \partial_h (v_2 \Gamma_h) - \Gamma_h \partial_h v_2 = \nabla \cdot (v_2 \Gamma_h \mathbf{s}_h) - \Gamma_h \partial_h v_2 
\end{align}
Utilizing the divergence theorem, we can transform the volume integrals over the divergence terms into the following surface integrals:
\begin{align}
\int_\Omega dV \ \nabla \cdot (v_1 \Gamma_0 \mathbf{s}_0)
& =  \int_{\delta \Omega} dS \ v_1 \Gamma_0 
 \mathbf{s}_0  \cdot \mathbf{n} \\ 
\int_\Omega dV \ \nabla \cdot (v_2 \Gamma_h \mathbf{s}_h)
& =  \int_{\delta \Omega} dS \ v_2 \Gamma_h 
 \mathbf{s}_h  \cdot \mathbf{n},
\end{align}
where $\mathbf{n}$ is the unit outward normal on $\delta \Omega$. Thus we finally obtain 
\begin{subnumcases}{}
\int_\Omega dV \ \left( \Gamma_0 \partial_0 v_1
+i c_0 v_1\Gamma_0  + i c_{\bar{h}} v_1\Gamma_h \right)
-  \int_{\delta \Omega} dS \ v_1 \Gamma_0 
 \mathbf{s}_0  \cdot \mathbf{n} = 0 \label{eq:weakformTTa}
 \\ 
\int_\Omega dV \ \left( \Gamma_h \partial_h v_2
+i c_0' v_2\Gamma_h  + i c_{h} v_2\Gamma_0 \right)
-  \int_{\delta \Omega} dS \ v_2 \Gamma_h 
 \mathbf{s}_h  \cdot \mathbf{n} = 0 \label{eq:weakformTTb}
\end{subnumcases}
where $c_0' = c_0 + \partial_h (\mathbf{h}\cdot\mathbf{u})$.

Equations~\eqref{eq:weakformTTa} and \eqref{eq:weakformTTb} represent the so-called \emph{weak} or \emph{variational formulation} of the differential equations \eqref{eq:TT_gammaa} and \eqref{eq:TT_gammab}. The test functions $v_1$ and $v_2$ as well as $\Gamma_0$ and $\Gamma_h$ are assumed to belong to an infinite dimensional Hilbert space $H$. It is required that these equalities hold for all test functions in $H$.
In practice, however, the application of FEM on these functions converts them to functions in a finite dimensional function space and then in ordinary Euclidean vectors (in a vector space) that can be managed via numerical methods.

This formulation is called \emph{weak} because it relaxes the requirement expressed by \eqref{eq:TT_gammaa} and \eqref{eq:TT_gammab}, where all the terms of the PDE must be defined in each point (pointwise formulation).  The relations in \eqref{eq:weakformTTa} and \eqref{eq:weakformTTb}, instead, only entail equality in an integral sense. As an example, a first derivative discontinuity of the solution function does not preclude integration. It introduces, however, a
distribution (in mathematical sense) for the second derivative. It is important to notice that in such a case \eqref{eq:TT_gammaa} and \eqref{eq:TT_gammab} become immaterial in a discontinuity point.

In contrast to \eqref{eq:integralTTa} and \eqref{eq:integralTTb}, Equations \eqref{eq:weakformTTa} and \eqref{eq:weakformTTb} do not contain derivatives of the functions $\Gamma_0$ and $\Gamma_h$. They can be implemented in
a FEM code, using a mesh of two-dimensional elements (often triangles, but also rectangles or even higher order elements are used) adapted to the crystal shape in a quite straightforward fashion. 

The solution of \eqref{eq:weakformTTa} and \eqref{eq:weakformTTb} is expressed as $\Gamma_0(\mathbf{r}) = \sum_i \Gamma_{0,i} N_i(\mathbf{r})$ and
$\Gamma_h(\mathbf{r}) = \sum_i \Gamma_{h,i} N_i(\mathbf{r})$, where the sums go over all $n$ knots in the mesh, and $\Gamma_{0,i}$ and $\Gamma_{h,i}$ are coefficients to be determined. $N_i(\mathbf{r})$ are the basis (or shape) functions related to the $i$-th knot. Basis functions are non-zero everywhere except in the vicinity of the knot they are tied to. Customarily, they are polynomial (e.g. B-splines) functions of degree one or higher: in this work quadratic functions were used. The well-known Galerkin method (used also in this work) uses a set of test functions identical to the basis functions, \emph{i.e.} $v_{1,j}(\mathbf{r})=v_{2,j}(\mathbf{r})=N_j(\mathbf{r})$. 
By transforming Eqs.~\eqref{eq:TT_gammaa} and \eqref{eq:TT_gammab} into their weak form, the problem of solving this pair of PDEs is then reduced into solving a system of $2n$ algebraic linear equations from which the coefficients $\Gamma_{0,i}$ and $\Gamma_{h,i}$ are to be determined numerically.

One of the most outstanding assets of FEM is
its ability to choose test and basis functions among a wide host of functions. It is often beneficial to
select test and basis functions with a locally variable geometrical support. It should be reminded that
all the highlighted features reported above are not present in FD, thus making the FD solution of the same problem by far more laborious and less efficient than the analogous FEM solution.

\section{Notes on the reference implementation}

The method was implemented on a commercial modeling and simulation software COMSOL Multiphysics 5.3. COMSOL Multiphysics was chosen due to its widespread use and readily available structural mechanics and heat conduction modules that can be used to solve the deformation field for the TTE computation in the future development of the method. 

The method was implemented using the \emph{Weak Form PDE} interface which allows the user to easily include arbitrary weakly formulated differential equations into the system. In addition to the strain included in the TTE, the deformation was taken into account by including the displacement vector field $\mathbf{u}$ into the mesh geometry through the \emph{Moving Mesh} interface. 

Meshing of the crystal domain was done using the \emph{Free Triangular} node which automatically generates an unstructured mesh grid of triangular elements according to the given limitations on element sizes etc. The grid parameter of most relevance to this work was the \emph{Maximum element size} which limits the maximum distance of the grid nodes. For a simple rectangular geometry this value corresponds to the typical node separation inside the domain. Later in the text we refer to this parameter simply as the (grid) element size. 

\begin{figure}
\label{fig:vectors}
\centering
\includegraphics[width=0.7\textwidth]{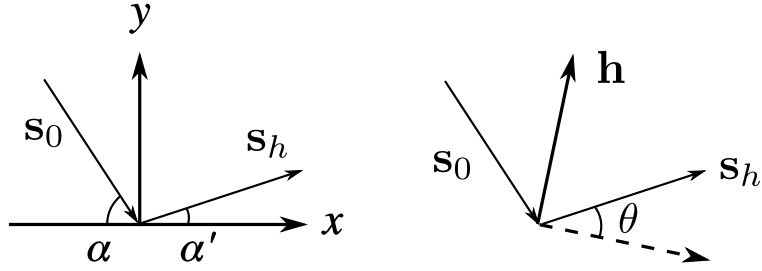}
\caption{Direction vectors $\mathbf{s}_0$ and $\mathbf{s}_h$ of incident and diffracted
wave with respect to the used Cartesian coordinate system $(x,y)$. The sign convention is so that both angles are positive in the case depicted in the figure. Note that the coordinate system is not in the general case aligned with $\mathbf{h}$.}
\end{figure}

As COMSOL uses a Cartesian coordinate system, the oblique coordinates $(s_0,s_h)$ need to transformed into Cartesian ones. The relation between the two systems is presented in Figure~\ref{fig:vectors}. The unit vectors $\mathbf{s}_0$ and
$\mathbf{s}_h$ in the Cartesian basis $\{\mathbf{e}_x, \mathbf{e}_y\}$ are
\begin{equation}
\mathbf{s}_0 = \cos \alpha \mathbf{e}_x - \sin \alpha \mathbf{e}_y \qquad \mathbf{s}_h = \cos \alpha' \mathbf{e}_x + \sin \alpha' \mathbf{e}_y.
\end{equation} 
Thus the oblique coordinates in terms of $x$ and $y$ are 
\begin{equation}
s_0 = \frac{x \sin \alpha' - y \cos \alpha'}{\sin(\alpha+\alpha')} \qquad
s_h = \frac{x \sin \alpha + y \cos \alpha}{\sin(\alpha+\alpha')},
\end{equation}
and the partial derivatives
\begin{equation}
\frac{\partial}{\partial s_0} =
\cos \alpha \frac{\partial}{\partial x}  - \sin \alpha \frac{\partial}{\partial y} \qquad
\frac{\partial}{\partial s_h} =
\cos \alpha' \frac{\partial}{\partial x}  + \sin \alpha' \frac{\partial}{\partial y} 
\end{equation}

For the propagation calculations, the solved complex wave amplitudes on the crystal surface are exported into a text file. The contents of files are read and the Fresnel integral is calculated for all the points on the detector plane using a Python program. 

\section{Simulations}

In order to validate our FEM method, we solved the TTE for the symmetric Si(111) reflection for an undeformed and a cylindrically bent crystal for various incidence angle. The energy of the $\sigma$-polarized incident X-rays was set to $E=6$~keV. The $\chi_{0,h,\bar{h}}$ values together with diffraction related quantities were computed with XOP 2.4 \cite{sanchezdelrio11,sanchezdelrio15} and are presented in Table \ref{tbl:crystaldata}.

\begin{table}[]
\centering
\caption{Crystal parameters and diffraction related quantities for symmetric Si(111) at photon energy of 6 keV  ($\sigma$-polarized).}
\label{tbl:crystaldata}
\begin{tabular}{rl}
\multicolumn{1}{l}{}                        &                                                       \\
\multicolumn{1}{r|}{$\chi_0$}               & $-0.274564 \cdot 10^{-4} + i 0.109657 \cdot 10^{-5}$  \\
\multicolumn{1}{r|}{$\chi_h$}               & $-0.109980 \cdot 10^{-4} - i 0.991441 \cdot 10^{-5}$  \\
\multicolumn{1}{r|}{$\chi_{\bar{h}}$}       & $-0.991441 \cdot 10^{-5} + i  0.109980 \cdot 10^{-4}$ \\
\multicolumn{1}{r|}{Bragg angle $\theta_B$} & 19.24$^\circ$                                         \\
\multicolumn{1}{r|}{Interplanar distance}   & 3.14 \AA                                              \\
\multicolumn{1}{r|}{Absorption length}      & 29.99~$\upmu$m                                         \\
\multicolumn{1}{r|}{Darwin width}           & 9.83 arcsec (47.7 $\upmu$rad)                                           \\
\multicolumn{1}{r|}{Extinction depth}      & 0.73~$\upmu$m                                         \\
\multicolumn{1}{r|}{Extinction length}      & 2.22~$\upmu$m                                         
\\
\multicolumn{1}{r|}{Refraction correction}      & 9.12~arcsec (44.2 $\upmu$rad)                                          
\end{tabular}
\end{table}

\subsection{Reflectivity curves of undeformed crystal}

The TTE were solved for various incidence angles for an undeformed, rectangular crystal slab. The thickness $t$ of the crystal was set to 50 $\upmu$m. In order to avoid the disturbances due to the sides of the crystal, the incident plane wave \eqref{eq:planewave} was multiplied with a Gaussian window function. The full width at half maximum (FWHM) of the window was chosen to be 100 $\upmu$m. It should be noted that this FWHM applies for the \emph{amplitude} of the wave; for the \emph{intensity}, the given value should be regarded as the full width at the quater of the maximum (FWQM). The width of the crystal was chosen to be 200 $\upmu$m which accommodates the masked beam well. For the aforementioned parameters, the simulated crystal can be considered thick in terms of the diffraction.
 
\begin{figure}
\label{fig:unbent_rocking_curves}
\centering
\includegraphics[width=\textwidth]{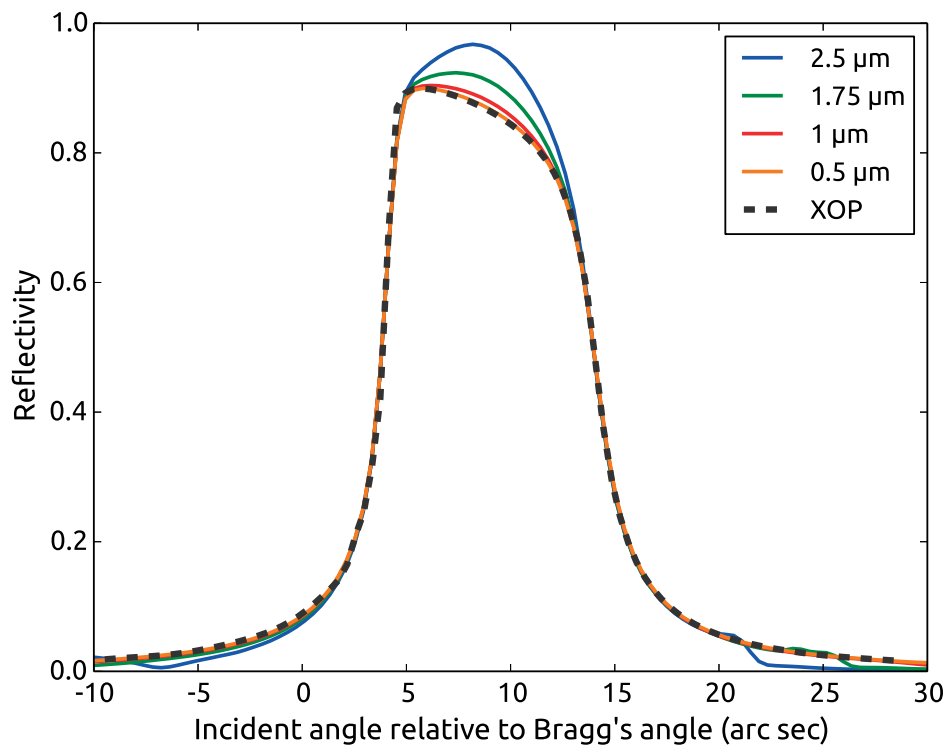}
\caption{The reflectivity curves of an undeformed Si(111) crystal computed with different grid element sizes. The reference curve is obtained using XINPRO in XOP.}
\end{figure}

The reflectivity or rocking curves (RC) were solved using maximum triangular element sizes of 0.5, 1, 1.75, and 2.5~$\upmu$m (Fig.~\ref{fig:unbent_rocking_curves}). As expected, the result converges towards the reference curve computed with XOP as the element size gets smaller.
The largest effect of the grid density can be seen on the top of the curve. This is natural since this is the region of the rocking curve where the dominating length scale is the extinction length. Comparing to the extinction length of 2.22~$\upmu$m, we find that the grid size needs to be 2-3 times smaller in order to get satisfactory convergence.

The deviations seen in the tails of RC for 1.75, and 2.5~$\upmu$m grids arise from the oscillating phase factor $\exp(-i\beta s_h)$ in the boundary condition of $\Gamma_0$. Using Bragg's law, the phase factor on the top surface can be written as 
\begin{equation}
\exp(-i \beta s_h) = \exp \left(-i k \Delta \theta \sin 2 \theta_B  \frac{x}{2 \cos \theta_B} \right) = \exp \left(-2 \pi i  \frac{\Delta \theta}{2 d} x \right),
\end{equation}
where $d$ is the interplanar distance of the Bragg planes. According to the Nyquist–Shannon sampling theorem, in order to sample a function with the frequency of $\Delta \theta/2 d$, the separation of grid points $\Delta x$ needs to be $\leq d/|\Delta \theta|$. For $d=3.14$~\AA \ and $\Delta \theta = 30$~arcsec (145.4 $\upmu$rad), we find that $\Delta x \leq 2.2$~$\upmu$m, which is in accordance with the simulations.

\begin{figure}
\label{fig:boundary_layer}
\centering
\includegraphics[width=0.4\textwidth]{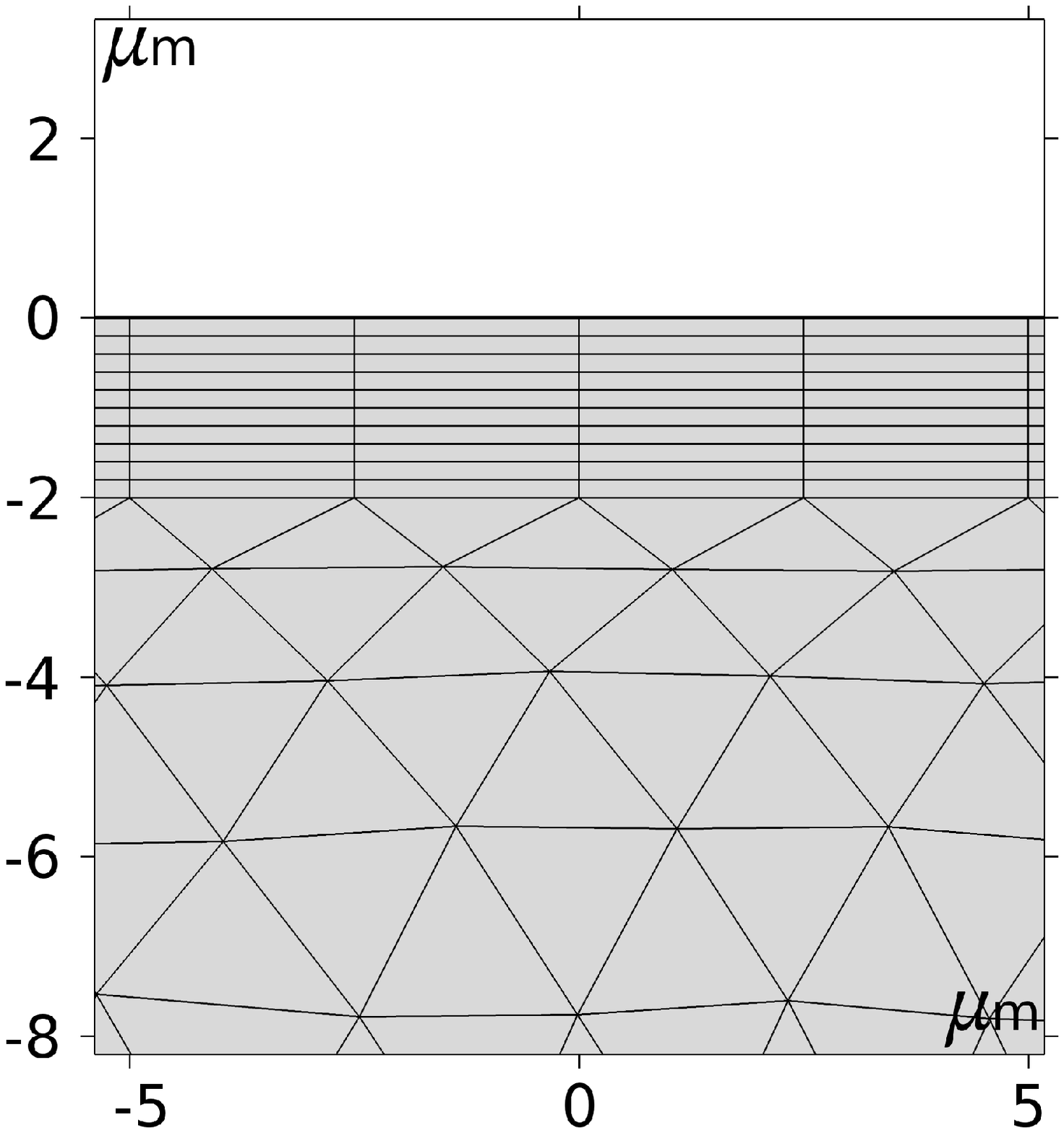}
\includegraphics[width=0.58\textwidth]{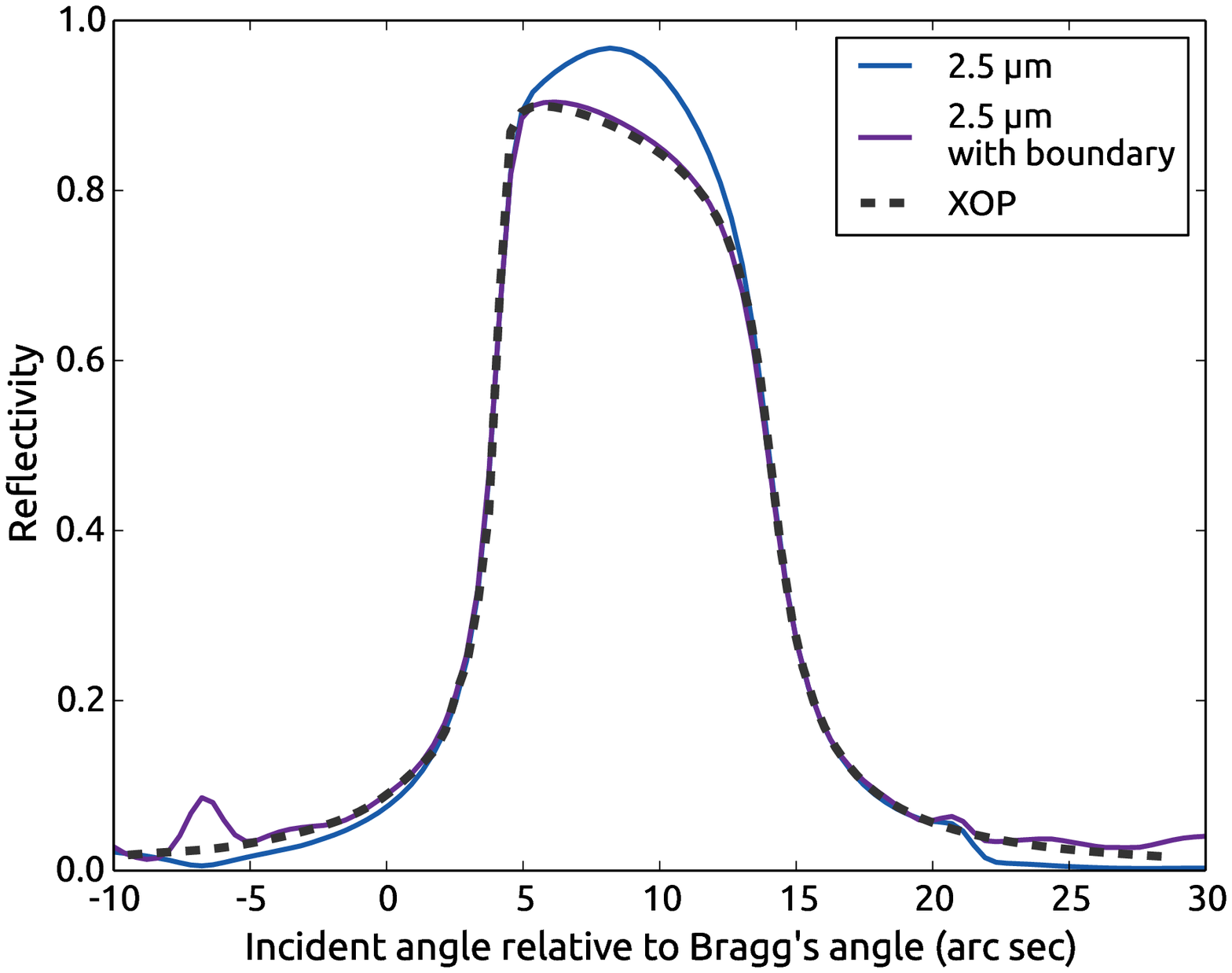}
\caption{\emph{LEFT:} Dense boundary layer on the surface. The number of rows of elements was 10 and the separation between rows was 0.2~$\upmu$m; \emph{RIGHT:}
The reflectivity curves computed on 2.5~$\upmu$m grid with and without the boundary layer.}
\end{figure}

A natural choice in the FEM framework would be to modify the mesh density locally, following the variations in the length scale of the solution.
For the problem at hand, the shorter length scale variation (extinction length) takes place near the surface. Using the \emph{Boundary Layer} option available in COMSOL, one can increase the density of the mesh near the surface and thus improve the accuracy of the RC (Fig.~\ref{fig:boundary_layer}). However, the introduction of the boundary layer causes some deviations in the tails of the RC, raising the need for a deeper investigation of the meshing process, which is outside the scope of the present work.

%\begin{figure}
%\label{fig:unbent_wavefield}
%\centering
%\includegraphics[width=\textwidth]{unbent_wavefield}
%\caption{asdasdadssads}
%\end{figure}

\subsection{Reflectivity curves of cylindrically bent crystal\label{sec:refl_cyl}}

The effect of cylindrical bending on the reflectivity properties was investigated by including the bending field presented in Appendix \ref{app:deformation}. Using the Poisson ratio $\nu=0.27$ and the crystal parameters tabulated in Table~\ref{tbl:crystaldata}, the cylindrical bending can be considered weak when the bending radius $R \gg 0.12$~m, according to Eq. \eqref{eq:weak_bending}.

To examine the weakly deformed case, we set $R=5$~m. 
The crystal thickness was 50~$\upmu$m, as before, but the width was extended by adding 200~$\upmu$m to the right hand side (totalling 500~$\upmu$m) in order to accommodate the curved beam path. A line source boundary condition \eqref{eq:pointsourcewave} multiplied with a Gaussian window (FWHM of amplitude = 100 $\upmu$m) was used for the incident wave. The grid element size of 0.75~$\upmu$m was used. 

\begin{figure}
\label{fig:5m_bent_rocking_curves}
\centering
\includegraphics[width=\textwidth]{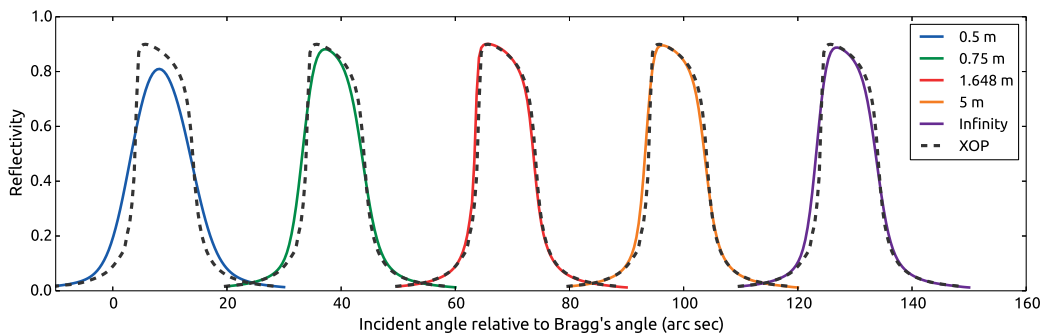}
\caption{The reflectivity curves of an cylindrically bent ($R = 5 m$) Si(111) crystal computed for various source distances using grid element size of 0.75~$\upmu$m. The source distance $p=1.648$~m is the on-Rowland circle position $p=R \sin \theta_B$. The curves are shifted on x-axis for clarity. The reference curve is the same as in Fig.~\ref{fig:unbent_rocking_curves}.}
\end{figure}

\begin{figure}
\label{fig:mirage}
\centering
\includegraphics[width=\textwidth]{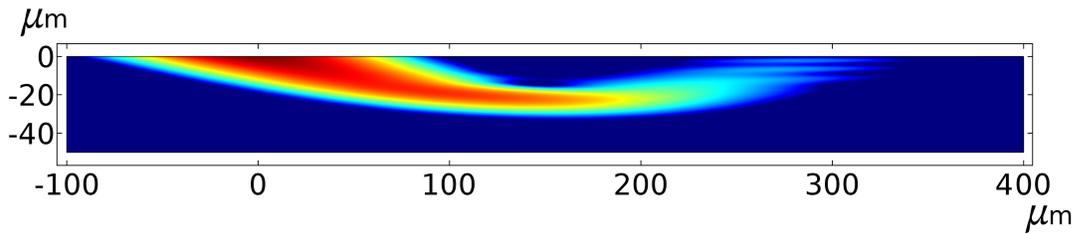}
\caption{Total intensity inside the crystal at $\Delta \theta = 1.31$ arc sec (6.37 $\upmu$rad) illustrating the mirage effect. Logarithmic scale is used for color mapping for visual clarity.}
\end{figure}

The RCs for various source distances are presented in Figure~\ref{fig:5m_bent_rocking_curves}. When the line source is on the Rowland circle ($p = 1.648$~m), the RC deviates only little from the undeformed reference curve which is an expected result for a weakly deformed crystal. The slight shift of the curve to the left can be associated with increased lattice spacing on the upper part of the crystal owing to the bending and  the non-zero Poisson ratio. A part of the added weight on the left side can also be explained by the so-called mirage effect \cite{gronkowski84, gronkowski91,authierbook} which can be seen in Figure~\ref{fig:mirage} where the total intensity inside the crystal at $\Delta \theta = 1.31$ arc sec (6.37 $\upmu$rad) is visualized. The incident cylindrical wave, which is approximately a plane wave with the incidence angle outside the Darwin range\footnote{Taking into account the refraction correction, the Darwin range is 4.21 arcsec $< \Delta \theta <$ 14.03 arcsec (or 20.4 $\upmu$rad $< \Delta \theta <$ 68.0 $\upmu$rad)}, excites a wavefield deep in the crystal due to the locally changing orientation of the reflecting planes along the beam trajectory in the crystal.

When the distance of the source is changed, we observe that RCs broaden. This is expected as the incidence angle is not the same at every point of the crystal surface if the source is taken off the Rowland circle and thus the diffraction condition is not fulfilled on the whole incident wavefront.

\begin{figure}
\label{fig:half_m_bent_rc_convergence}
\centering
\includegraphics[width=\textwidth]{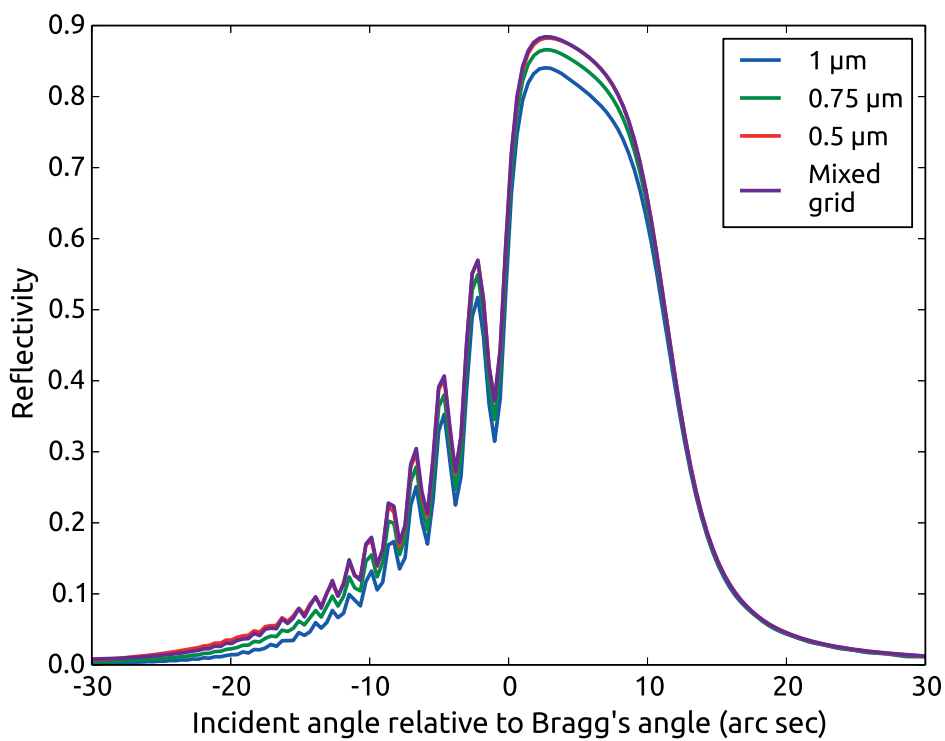}
\caption{The reflectivity curves of an cylindrically bent ($R = 0.5 m$) Si(111) crystal computed for various mesh grid element sizes. The source was on the Rowland circle ($R \sin \theta_B = 0.165$~m) and FWHM of the Gaussian window was 100~$\upmu$m.}
\end{figure}

The method was also tested for a smaller bending radius. To reflect the contemporary state of bent crystal analyser technology \cite{rovezzi17},
$R=0.5$~m was chosen. As for the unbent and 5~m bent cases, the method was found to be stable even when the shape of the RC is considerably affected by the deformation field. However, the convergence was found to be slower than previously, requiring grid element sizes of 0.5~$\upmu$m or even smaller. It turns out that the requirement for the increased grid density arises mainly from the stronger deformation in $y$-direction. By taking the advantage of the freedom in the mesh construction in the FEM scheme, we scaled the distance of the elements by factor of 0.5 in $y$-direction leading to a grid with mixed element dimensions of $0.5$~$\upmu$m$\times$1.0~$\upmu$m in vertical and horizontal directions, respectively. The resulting curve was found to be the same as with the uniform $0.5$~$\upmu$m$\times$0.5~$\upmu$m mesh but with half the amount of grid elements,
which again echoes the benefits of adaptive grid construction schemes for speed and memory optimization. The RCs computed with different grids are presented in Figure~\ref{fig:half_m_bent_rc_convergence}.

\begin{figure}
\label{fig:half_m_bent_rc_beamsize}
\centering
\includegraphics[width=\textwidth]{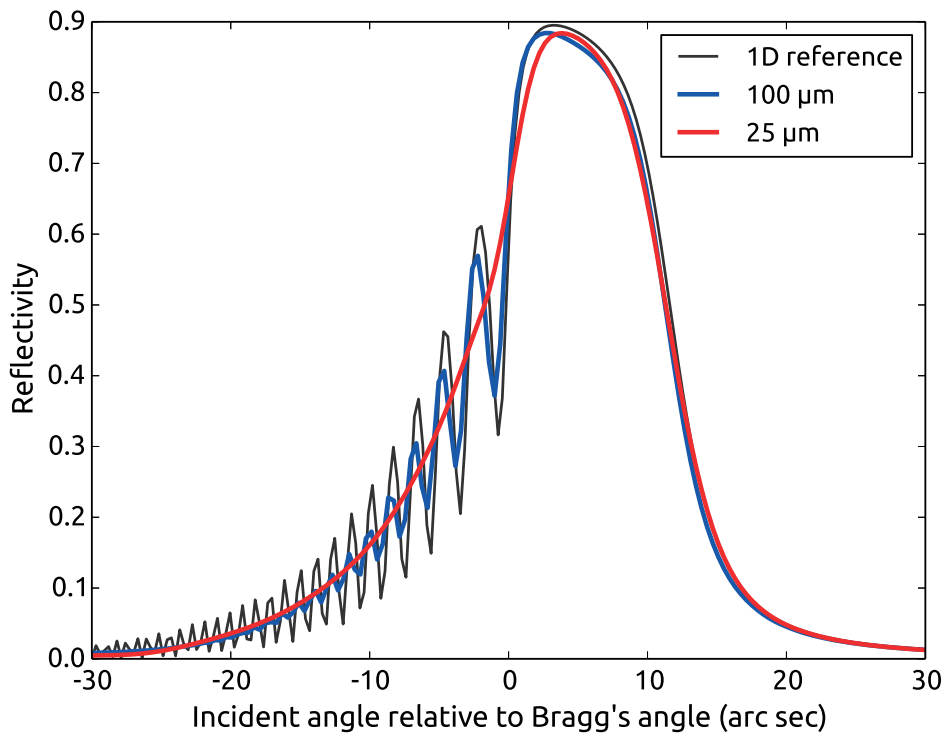}
\caption{The reflectivity curves of an cylindrically bent ($R = 0.5 m$) Si(111) crystal computed for two differently sized Gaussian windows (25~ and 100~$\upmu$m). The source is on the Rowland circle ($R \sin \theta_B = 0.165$~m). The reference curve is the solution to 1D form of Takagi-Taupin equations.}
\end{figure}

In order to validate the method for $R=0.5$~m, we computed the RC with a 1D Takagi-Taupin solver\footnote{A slightly modified version of \textsc{https://github.com/aripekka/pytakagitaupin}} and compared the mixed grid solutions to it. As shown in Figure~\ref{fig:half_m_bent_rc_beamsize}, the FEM results follow the general features of the reference curve but do not reproduce the the very same details. This is to be expected as the 2D situation allows for the lateral dependence in the TTE that is missing in the 1D case. This can be demonstrated by varying the footprint of the incident beam; For a sufficiently small beam the overlap between the incident and the diffracted waves is smaller, meaning that they interfere less with each other and thus lead to suppression of Pendell\"{o}sung oscillations. Fig.~\ref{fig:half_m_bent_rc_beamsize} is a great example of how the 1D TTE is inherently different from the true 2D solution. 

\subsection{Propagation and focusing of the diffracted beam}

\begin{figure}
\label{fig:intensity_phase_on_surf}
\centering
\includegraphics[width=0.49\textwidth]{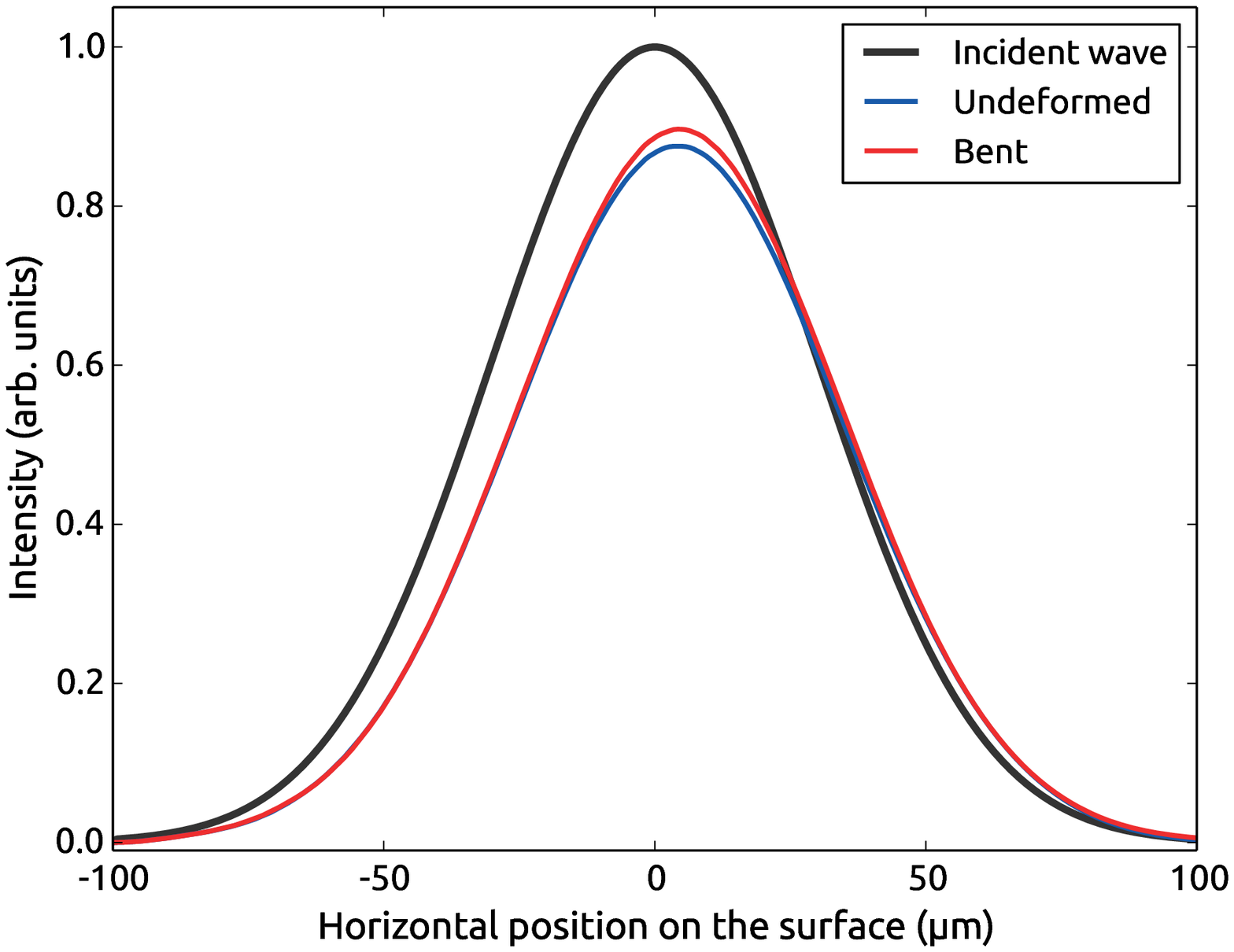}
\includegraphics[width=0.49\textwidth]{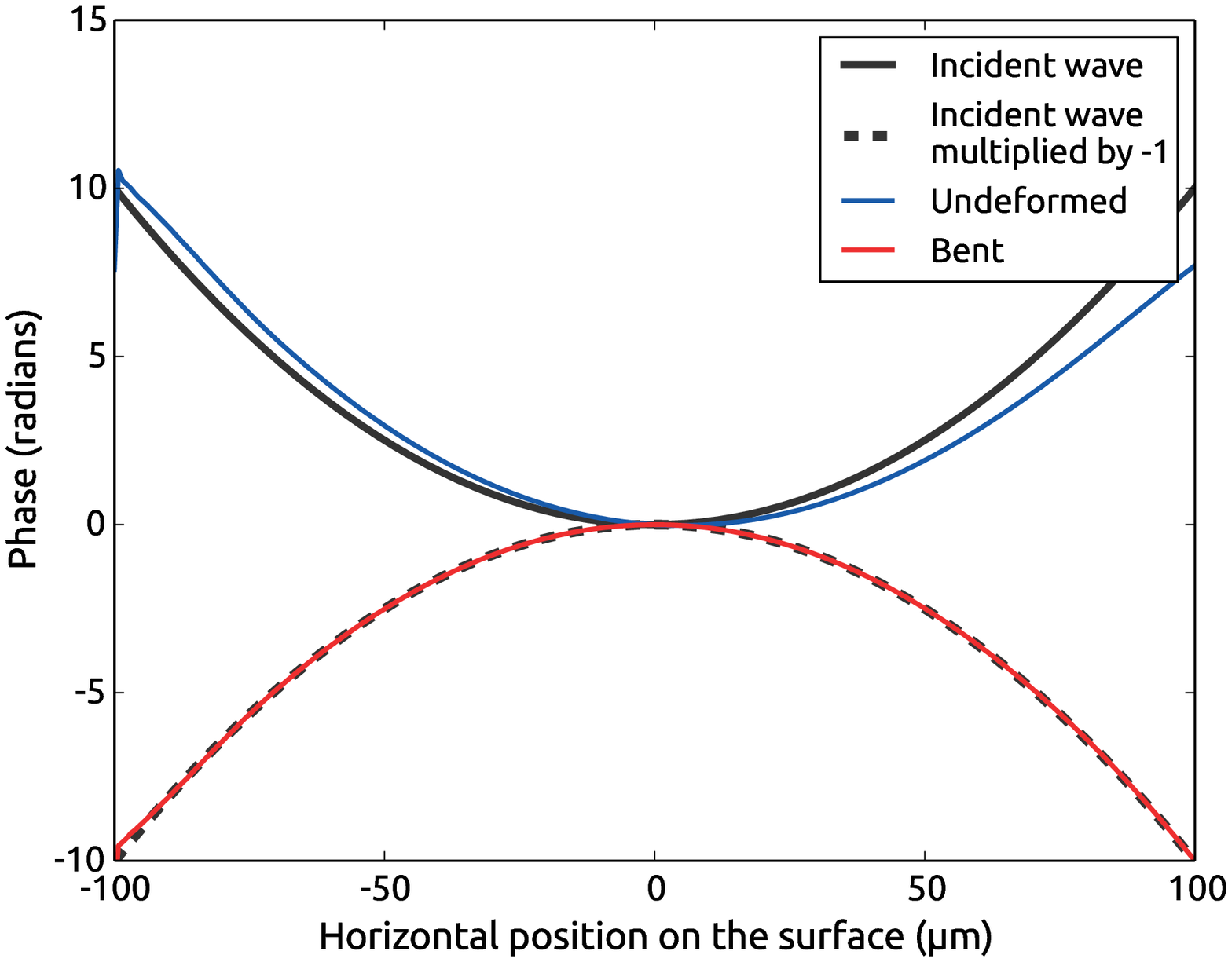}
\caption{LEFT: The intensity distribution of the incident and diffracted waves on the crystal surface for the 5~m cylindrically bent crystal and the undeformed crystal (FWQM of the incident curve is 100~$\upmu$m). The rocking angle was $\Delta \theta = 6.9697$ arc sec (33.79 $\upmu$rad) The point source was on the Rowland circle ($R \sin \theta_B = 1.648$~m). RIGHT: The phases of the same waves}
\end{figure}

We examine the propagation of the diffracted wave by the 5~m cylindrically bent crystal from section $\ref{sec:refl_cyl}$. The grid element size of 0.75~$\upmu$m was used. The diffracted intensity and the phase of the diffracted wave (without the plane wave factor $\exp(i\mathbf{k}_h\cdot\mathbf{r})$) on the crystal surface are presented in Figure~\ref{fig:intensity_phase_on_surf} on the top of the diffraction curve at $\Delta \theta = 6.9697$~arc~sec (33.79~$\upmu$rad). In addition to the bent crystal, the curves for the similar but undeformed crystal are shown for comparison. 

The intensity distribution is found to be similar for both bent and undeformed crystals. This is not surprising as the reflectivity curve is fairly unaffected by the deformation field. However, the phases of the diffracted waves differ drastically, owing mainly to the deformation phase factor 
$\exp(-i\mathbf{h}\cdot\mathbf{u})$. One should expect such a difference as the proper focusing in the Rowland circle geometry is dependent on the correct bending of the crystal. The phase of the diffracted wave by the bent crystal coincides with the phase of the incident wave multiplied by -1, which indicates that it describes a spherical wave with the same focal distance as the incident wave but propagating \emph{to} the focal point instead of \emph{from} it.

\begin{figure}
\label{fig:focus_comparison}
\centering
\includegraphics[width=0.49\textwidth]{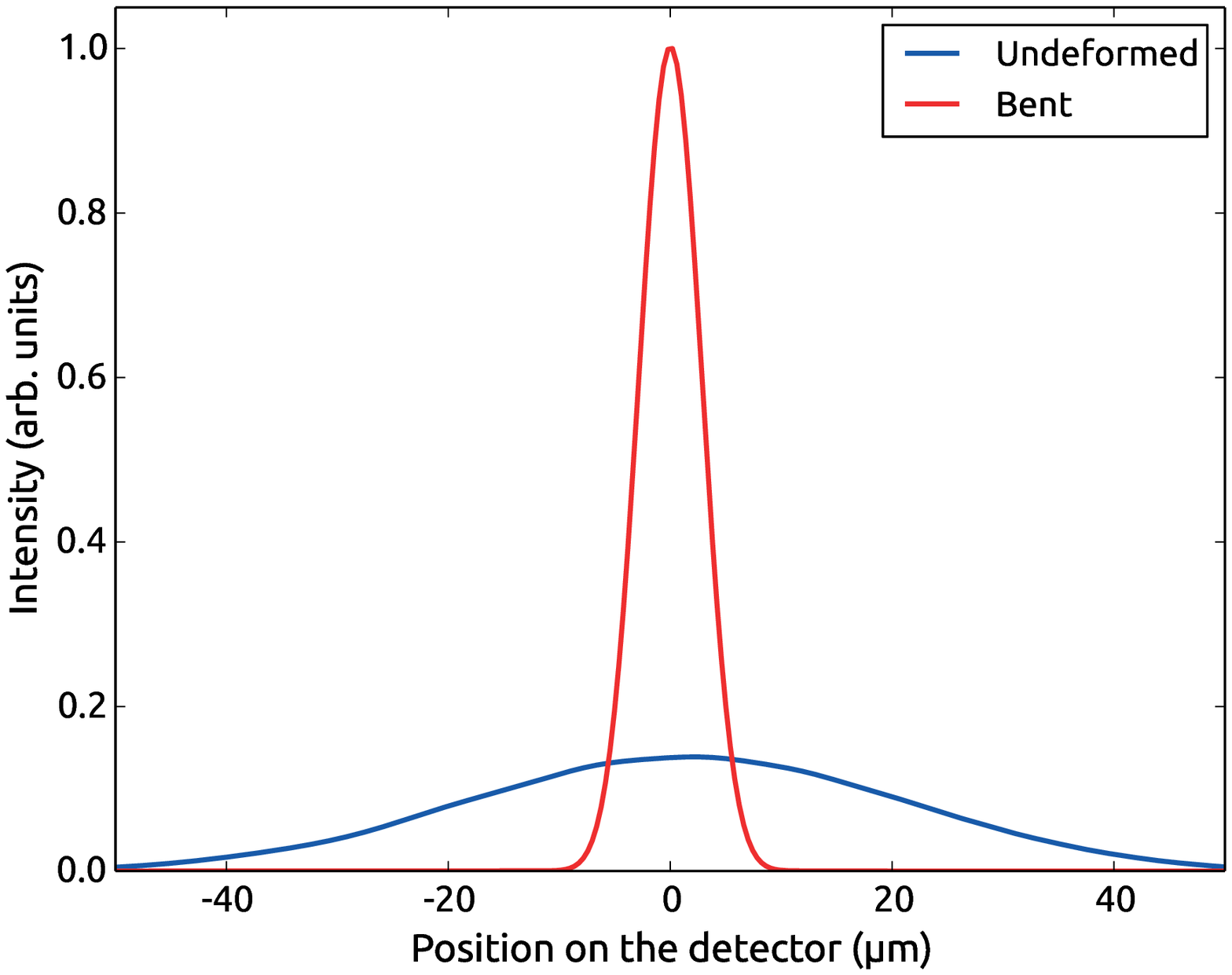}
\includegraphics[width=0.49\textwidth]{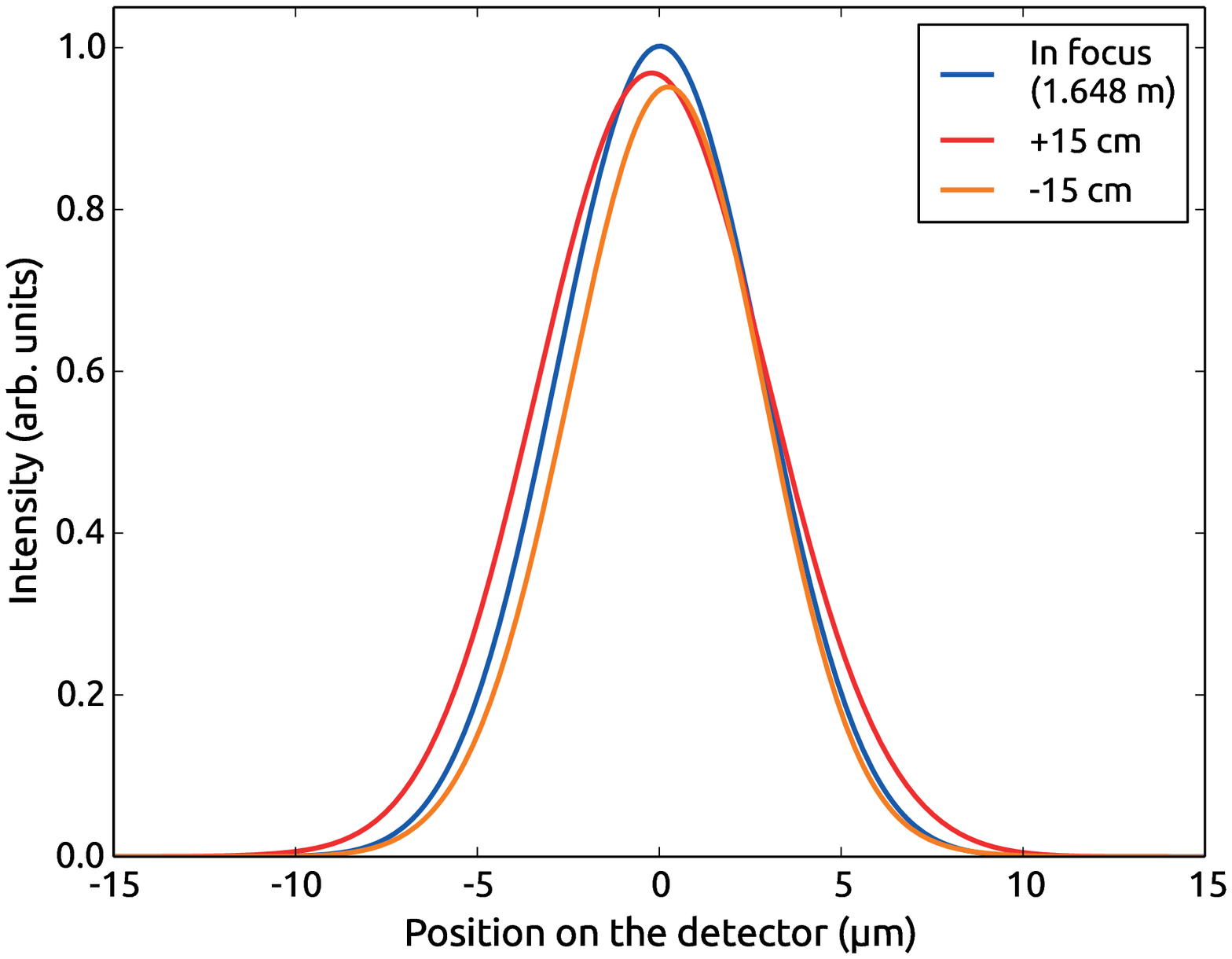}
\caption{LEFT: Computed intensity distribution at the focal position on the Rowland circle. RIGHT: Effect of the displacement of the detector through the focal spot of the cylindrical crystal.}
\end{figure}

The proper focusing is confirmed by computing the Fresnel integral \eqref{eq:fresnel_integral} over the crystal surfaces. The intensities of the  propagated waves on the Rowland circle are presented in Figure~\ref{fig:focus_comparison}. We indeed observe that the bent crystal focuses the beam whereas for the unbent one the diffracted wave diverges. It is also confirmed that the optimal focal position of the cylindrical crystal is situated on the Rowland circle as the peak intensity of the focal point decreases when the distance of the detector plane is altered.

\begin{figure}
\label{fig:mirage_focusing}
\centering
\includegraphics[width=0.49\textwidth]{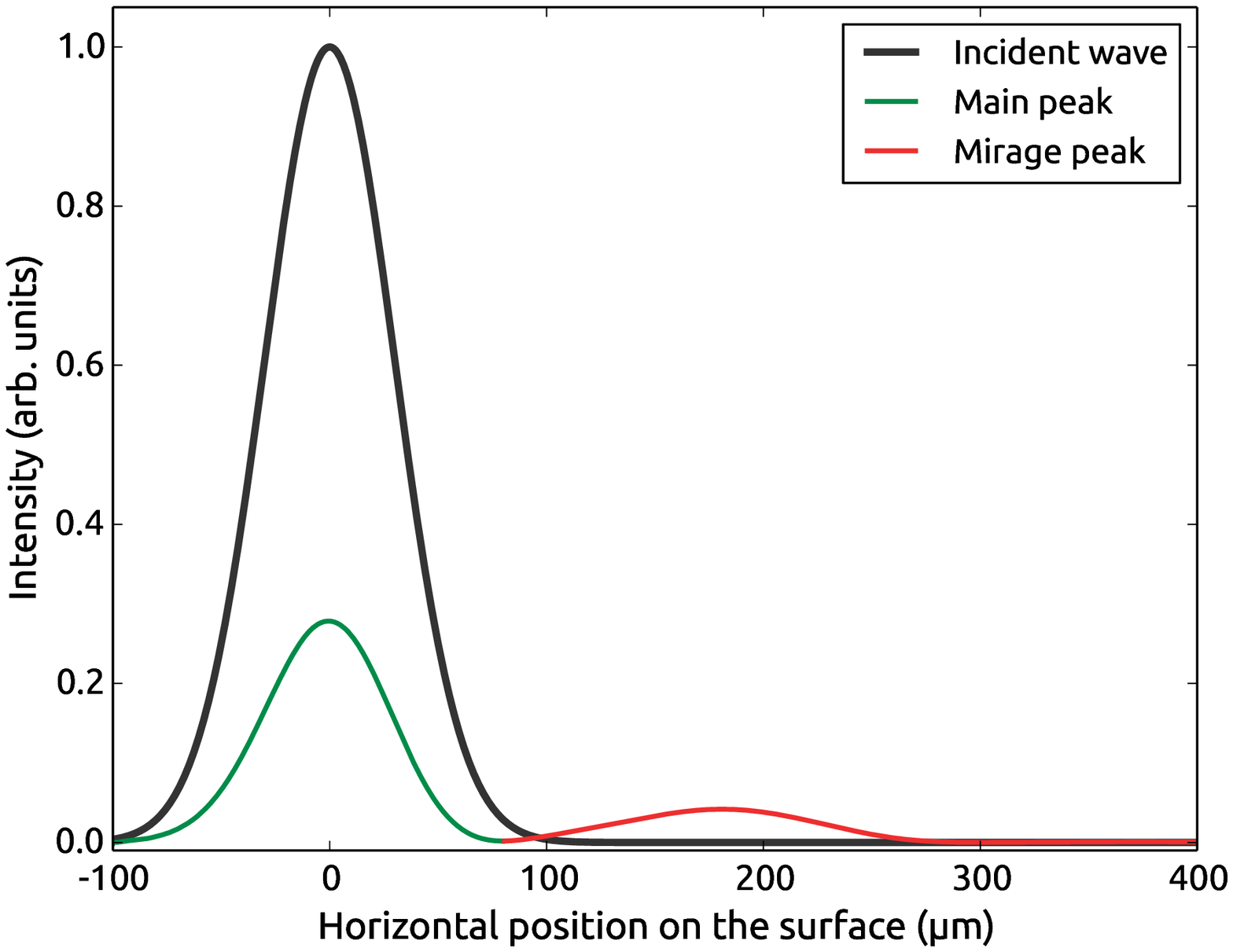}
\includegraphics[width=0.49\textwidth]{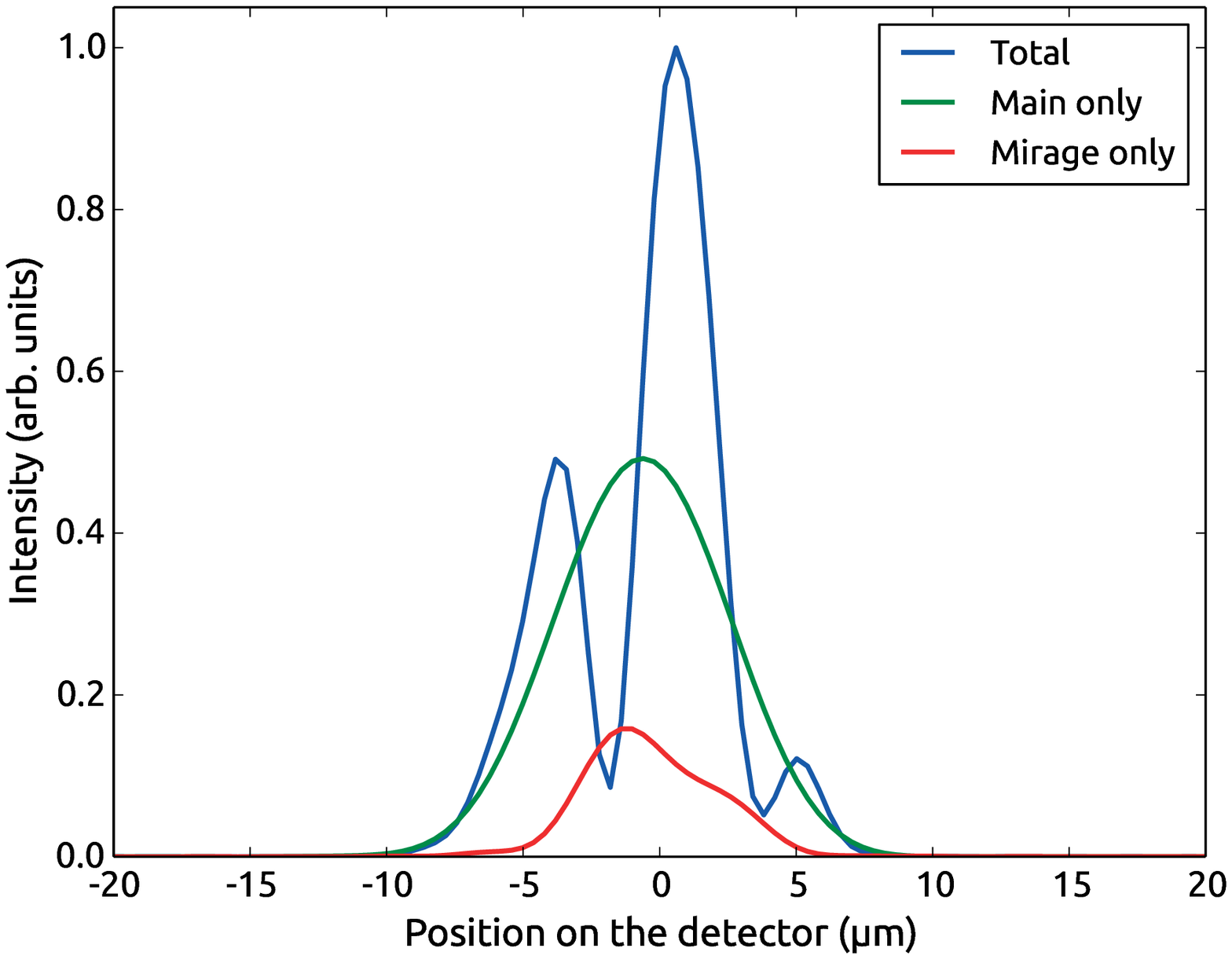}
\caption{LEFT: The intensity distribution of the incident and diffracted waves on the crystal surface for the 5~m cylindrically bent crystal showing the mirage effect (FWQM of the incident curve is 100~$\upmu$m). The rocking angle is $\Delta \theta = 2.9293$ arc sec (14.20 $\upmu$rad). The point source is on the Rowland circle ($R \sin \theta_B = 1.648$~m). RIGHT: Propagated intensities at detector plane on the Rowland circle. The structure in the total intensity arises from the phase differences between the main and the mirage peak.}
\end{figure}

The behaviour of the focus in the presence of the mirage peak was also investigated. Figure~\ref{fig:mirage_focusing} shows the intensities on the crystal surface and on the detector plane at the Rowland circle for $\Delta \theta = 2.9293$ arc sec (14.20 $\upmu$rad). In addition to computing the Fresnel integral over the whole surface, we divided the diffracted wave into the main peak and mirage peak at $x=80$~$\upmu$rad and propagated the peaks separately to the detector plane. As it can be seen, the peaks focus nicely on the Rowland circle when propagated separately, with the mirage peak showing a slight shoulder on the right side. However, when propagated together, they form a three-peaked structure in the intensity distribution. This interference related phenomenon is caused by the phase differences between the main peak and the mirage peaks which arise at different depths in the crystal. 

\begin{figure}
\label{fig:focal_distances}
\centering
\includegraphics[width=\textwidth]{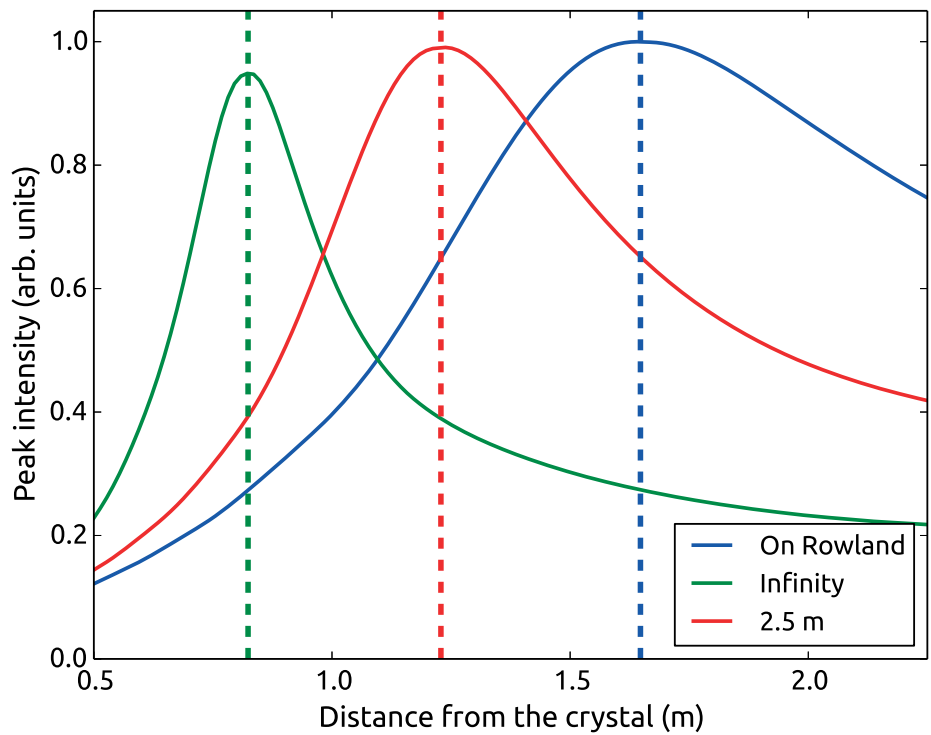}
\caption{The peak intensities of the propagated wave fields as a function of distance of the detector plane for various source distances. The dashed vertical lines correspond to the focal lengths calculated using Equation~\eqref{eq:lens_equation}. }
\end{figure}

Finally, the focal length for various source distances was studied. According to the lens equation \cite{chukhovskii92}, the source distance $d$ and the focal length $d$ are related by 
\begin{equation}
\frac{1}{p}+\frac{1}{q}=\frac{2}{R \sin \theta_B}\label{eq:lens_equation}.
\end{equation}
The validity of this relationship was investigated by computing the profiles of the propagated waves as a function of the detector plane distance for three different source distances. The peak intensities of the profiles were obtained and are plotted together with the predictions of Eq.~\eqref{eq:lens_equation} in Figure~\ref{fig:focal_distances}. The maxima of the peak intensities are found to be well in accordance with the lens equation, which is an expected result.

\section{Application to experiments}
For the sake of generality of our approach, all calculations in the previous sections were performed under ideal or simplified conditions: \emph{e.g.} monochromatic beam, no thermal load effects, perfectly cylindrical curvature, no mounting inaccuracies, and limited size of the incoming beam footprint on
the crystal surface. The investigation of all these effects would depend strongly on each individual beamline optical setup to be examined, and this would be far out of the scope of the present research.
However, the results of our simulations can be effectively used \emph{e.g.} to predict the lowest size limits of the focal spots produced by bent crystals in given optical arrangements.

Bearing in mind this point of view, we considered the example of a knife-edge scanning measurement carried out at the dispersive EXAFS beamline ID24 of the ESRF \cite{hagelstein95}. The aim of the experiment was to determine the focus width using the Si(111) reflection of a symmetrically cut
curved polychromator at a mean photon energy $E=7$~keV. The source was a secondary source (just downstream from a demagnification mirror), the size of which was evaluated to be between 40~and
45~$\upmu$m$^2$. The distance to the curved polychromator was about $p=30$~m and the focal distance was found to be $q=0.75$~m. The radius of curvature was estimated to be approximately $R=5.3$~m. The illuminated area on the crystal was about 15~cm. 

\begin{figure}
\label{fig:experimental_intensity}
\centering
\includegraphics[width=\textwidth]{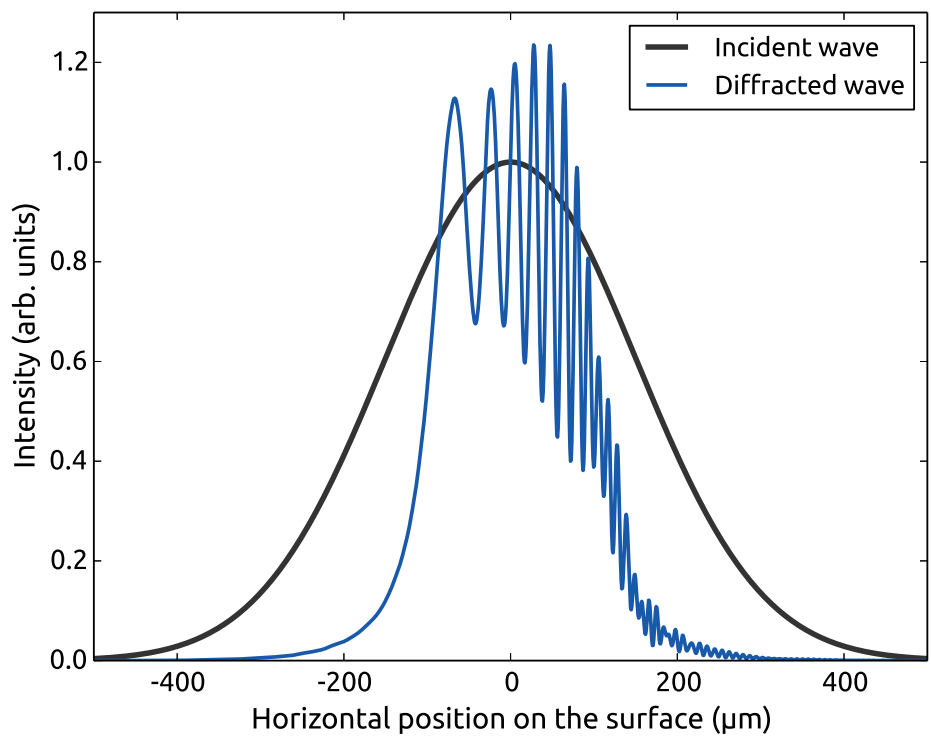}
\caption{Computed intensities of the incident and diffracted wave on the crystal surface for simulation parameters chosen to match the experimental conditions. The illuminated crystal surface is 1000~$\upmu$m wide and the FWQM of the incident intensity profile is 500~$\upmu$m. The width of the diffracted wave profile is roughly 200~$\upmu$m.}
\end{figure}

Following the simulation schemes described in the former sections (taking a point source) and using the same parameters as in the ID24 experiment, we first checked whether our simulations would be able to determine a focal distance matching the experimental one. The crystal dimensions of 1000~$\upmu$m (H) $\times$ 50~$\upmu$m (V) were used. The FWQM of incident intensity profile is 500~$\upmu$m. The incidence angle was chosen to be in the center of the rocking curve ($\Delta \theta = 7.65$ arc sec $=$ 37.1 $\upmu$rad) Since the source is not on the Rowland circle, much larger areas would not lead to a better result as the diffraction condition is not fulfilled further away from the crystal center. The simulated intensity profiles are presented in Figure~\ref{fig:experimental_intensity}. The oscillations in the diffracted profile are due to the interference between the incident and diffracted wave which became prominent when the source is not on the Rowland circle and the illuminated area is large enough.

\begin{figure}
\label{fig:experimental_focus}
\centering
\includegraphics[width=\textwidth]{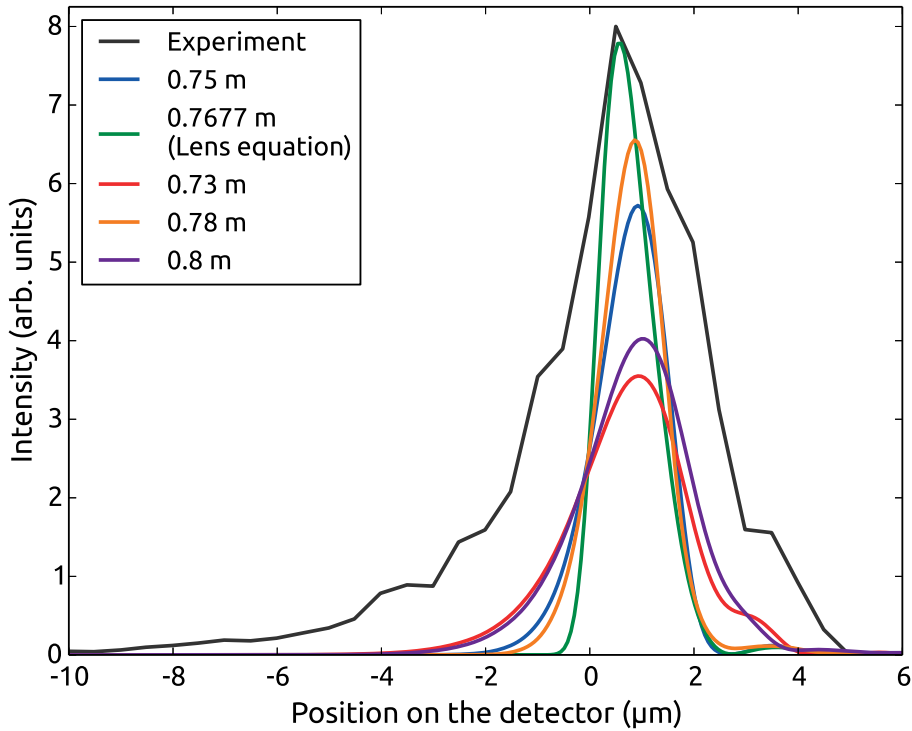}
\caption{Simulated focal profiles at various focal distances compared to the experimental focus.}
\end{figure}

The correctness of the prediction is demonstrated by Fig.\ref{fig:experimental_focus} showing focal spot distributions evaluated at different distances: the optimal focal distance is found to be 0.7677~m, in accordance with the Lens equation. The experimentally found value of $q=0.75$~m is not too far away from the simulations when taking into account that the FWHM of the focus does not vary significantly in the range of $\pm$0.02~m around the optimum. 

The calculated profile, as anticipated, is significantly narrower (Fig.\ref{fig:experimental_focus}): one can state that FWHM$_{exp}$ = 2.8~$\upmu$m, whereas FWHM$_{calc}$ = 1.4~$\upmu$m. The latter value could be thought of by the experimenter as an ideal target to assess the attained level of mounting accuracy of the setup. The small side peak on the left of the central one (Fig.\ref{fig:experimental_focus}) might be the signature of the intensity
spread due to the Johann error \cite{wang10}.

\section{Conclusions and outlook}

A general approach to solving numerically the Takagi-Taupin equations (TTE) in their integral representation via FEM, as implemented in the COMSOL Multiphysics package, is presented. This provides a wide flexibility in the numerical solution of dynamical diffraction problems for both perfect and deformed crystals of any arbitrary shape. The solution is computationally efficient and comparable to the classical albeit less flexible and powerful finite difference approach in conventional cases of simplistic geometries, i.e. 2D Cartesian systems with mostly straight line boundaries. In particular, we have shown the versatility of the FEM computational frame in elucidating a wealth of noteworthy, yet hitherto not utterly explained aspects of the focusing behaviour of cylindrically bent crystals in symmetric Bragg geometry.

The outreach of our approach is meant to go far beyond the limited scope of the present work in order to encompass a vast class of numerical problems (not only 2-D but also 3-D) related to solving X-ray and neutron dynamical diffraction problems in Bragg and Laue geometry based on the solution of the TTEs, which can be tackled only numerically. For instance, the FEM method may be applied to investigating numerically the possibility to obtain rocking curves of the Darwin type in Bragg geometry instead of the Ewald type by using crystal plates featuring grooves on their back surface, as suggested by \cite{freund14}. Since FEM is one of the most powerful methods known to successfully address boundary and/or initial value problems described by PDEs and the COMSOL Multiphysics package allows integrating easily user-defined equation systems into its kernel and to show promptly the results thanks to its built-in graphical facilities, it is hoped to disclose end encourage the application of this more general technique among the relevant scientific community.
To this purpose, we have devised an open access location of our COMSOL files. They can be freely downloaded from \textsc{https://github.com/aripekka/fem-takagi-taupin}.

     % Appendices appear after the main body of the text. They are prefixed by
     % a single \appendix declaration, and are then structured just like the
     % body text.

\appendix
\section{Deformation field of cylindrically bent crystal\label{app:deformation}}

According to the conventional elastic theory of thin crystal plates of thickness $t$ \cite{nesterets06}, the deformation field  $(\mathbf{u}_x , \mathbf{u}_y)$ for a cylindrically bent isotropic crystal is given by: 
\begin{equation}
u_x = -\frac{x}{R}\left(y + \frac{t}{2} \right) \qquad
u_y = \frac{1}{2R}\left[x^2 + \nu \left(y + \frac{t}{2} \right)^2 \right],
\end{equation}
where $R$ is the bending radius and $\nu$ is the Poisson ratio. Top surface of the crystal before the deformation is assumed to at $y=0$ and the bottom at $y=-t$.

In the case of symmetric Bragg case, one obtains
\begin{equation}
\mathbf{h}\cdot\mathbf{u} = \frac{h}{2R}\left[x^2 + 
\nu \left(y + \frac{t}{2} \right)^2 \right]
= \frac{\pi}{R d}\left[x^2 + 
\nu \left(y + \frac{t}{2} \right)^2 \right],
\end{equation}
where $d$ is the interplanar distance of the considered Bragg reflection. Thus the deformation term $\partial_h (\mathbf{h}\cdot\mathbf{u})$ in becomes 
\begin{equation}
\frac{\partial (\mathbf{h}\cdot\mathbf{u})}{\partial s_h} = \frac{\partial (\mathbf{h}\cdot\mathbf{u})}{\partial x} \cos \theta_B + \frac{\partial (\mathbf{h}\cdot\mathbf{u})}{\partial y} \sin \theta_B
= \frac{2\pi}{Rd} \left[x \cos \theta_B + \nu \left(y+\frac{t}{2} \right) \sin \theta_B \right]. 
\end{equation}
Following \cite{gronkowski91}, the deformation is considered weak for symmetric Bragg case if 
\begin{equation}\label{eq:weak_condition}
\left|\frac{\partial^2 (\mathbf{h}\cdot\mathbf{u})}{\partial s_0 \partial s_h}\right|  \ll \frac{\pi^2 C^2 \left|\chi_h \chi_{\bar{h}} \right|}{\lambda^2 \tan \theta_B},
\end{equation}
%\begin{equation}\label{eq:weak_condition}
%\left| \frac{\partial^2 (\mathbf{h}\cdot\mathbf{u})}{\partial s_0 \partial s_h} L_{ex}^2 \right| \ll 1,
%\end{equation}
%where $L_{ex} = \lambda/\mathrm{Re}(\sqrt{\chi_h \chi_{\bar{h}}})$ is the extinction distance. 
Since
\begin{equation}
\frac{\partial^2 (\mathbf{h}\cdot\mathbf{u})}{\partial s_0 \partial s_h}
= \frac{2 \pi}{Rd}\left(\cos^2 \theta_B - \nu \sin^2 \theta_B \right),
\end{equation}
the condition \eqref{eq:weak_condition} takes the form
%\begin{equation}
%\frac{2 \pi\lambda^2}{Rd}\frac{\left|\cos^2 \theta_B - \nu \sin^2 \theta_B \right|}{\mathrm{Re}(\sqrt{\chi_h \chi_{\bar{h}}})^2} \ll 1
%\end{equation}
\begin{equation}\label{eq:weak_bending}
R \gg \frac{8 d \sin^2 \theta_B \tan \theta_B}{\pi C^2 \left|\chi_h \chi_{\bar{h}} \right|}
\left|\cos^2 \theta_B - \nu \sin^2 \theta_B \right|
\end{equation}

\section{Derivatives in oblique and Cartesian coordinate systems\label{app:identities}}
Consider a arbitrary function $F=F(x,y)$. In terms of the Cartesian coordinate basis $\{\mathbf{e}_x, \mathbf{e}_y\}$, the unit vectors, we define a oblique system $(s_0,s_h)$ with the base vectors
\begin{equation}
\mathbf{s}_0 = \cos \alpha \mathbf{e}_x - \sin \alpha \mathbf{e}_y \qquad \mathbf{s}_h = \cos \alpha' \mathbf{e}_x + \sin \alpha' \mathbf{e}_y.
\end{equation} 
Thus
\begin{equation}
\mathbf{s}_0 \cdot \nabla F = \left(\cos \alpha \mathbf{e}_x - \sin \alpha \mathbf{e}_y \right)\left(\mathbf{e}_x \frac{\partial F}{\partial x} + \mathbf{e}_y \frac{\partial F}{\partial y}\right) = \cos \alpha  \frac{\partial F}{\partial x} 
 - \sin \alpha \frac{\partial F}{\partial y}.
\end{equation}
Also 
\begin{equation}
 \nabla \cdot  (\mathbf{s}_0 F) = \left(\mathbf{e}_x \frac{\partial }{\partial x} + \mathbf{e}_y \frac{\partial }{\partial y}\right) \left(F \cos \alpha \mathbf{e}_x - F \sin \alpha \mathbf{e}_y \right)= \cos \alpha  \frac{\partial F}{\partial x} 
 - \sin \alpha \frac{\partial F}{\partial y}.
\end{equation}
Switching to the oblique system, we can write $F(s_0,s_h)=F(x(s_0,s_h),y(s_0,s_h))$.
Since $x=(s_0 \cos \alpha + s_h \cos \alpha')$ and
$y=(- s_0 \sin \alpha + s_h \sin \alpha')$,
\begin{equation}
\frac{\partial F}{\partial s_0} = \frac{\partial x}{\partial s_0}\frac{\partial F}{\partial x} + \frac{\partial y}{\partial s_0}\frac{\partial F}{\partial y} = \cos \alpha \frac{\partial F}{\partial x}- \sin \alpha \frac{\partial F}{\partial y}.
\end{equation}
Pulling all together, we thus obtain a useful identity
\begin{equation}
\mathbf{s}_0 \cdot \nabla F = \nabla \cdot  (\mathbf{s}_0 F)
= \frac{\partial F}{\partial s_0}.
\end{equation}
Analogously for $s_h$
\begin{equation}
\mathbf{s}_h \cdot \nabla F = \nabla \cdot  (\mathbf{s}_h F)
= \frac{\partial F}{\partial s_h}.
\end{equation}

     %-------------------------------------------------------------------------
     % The back matter of the paper - acknowledgements and references
     %-------------------------------------------------------------------------

     % Acknowledgements come after the appendices

\ack{\textbf{Acknowledgements}}

We are greatly indebted to Olivier Mathon (ESRF, Grenoble) for the provision of the experimental
data reported in the text. Computing resources were provided by the CSC-IT center for science,
Finland. One of the authors (A.-P.H.) was financially supported by the Academy of Finland (grant 1295696) and the doctoral programme in Materials Research and Nanosciences (MATRENA) at the University of Helsinki.

\bibliographystyle{iucr}
\bibliography{references}

%\begin{references}
%\reference{Author, A. \& Author, B. (1984). \emph{Journal} \textbf{Vol}, 
%first page--last page.}
%\end{references}

     %-------------------------------------------------------------------------
     % TABLES AND FIGURES SHOULD BE INSERTED AFTER THE MAIN BODY OF THE TEXT
     %-------------------------------------------------------------------------

     % Simple tables should use the tabular environment according to this
     % model

\end{document}